\documentclass[10pt,prd,preprintnumbers,floatfix,aps,nofootinbib,showpacs,twocolumn,superscriptaddress,noshowpacs]{revtex4-2} 

\usepackage{epsfig}
\usepackage{url}
\usepackage{hyperref}

\usepackage[normalem]{ulem} 

\usepackage{latexsym}
\usepackage{epsfig}
\usepackage{amsmath}
\usepackage{amssymb}
\usepackage[normalem]{ulem}
\usepackage{wasysym}
\usepackage{graphicx}
\usepackage{verbatim}
\usepackage{enumerate,mdwlist}
\usepackage[titletoc]{appendix}
\usepackage{amsfonts}
\usepackage{fancyvrb} 
\usepackage{tikz} 
\usetikzlibrary{calc}
\usepackage[export]{adjustbox}
\usepackage{multirow}

\usepackage[normalem]{ulem}

\newcommand{\tr}{\mbox{tr}}

\newcommand{\ba}{\begin{eqnarray}}
\newcommand{\ea}{\end{eqnarray}}
\newcommand{\be}{\begin{equation}}
\newcommand{\ee}{\end{equation}}

\newcommand{\nn}{\nonumber}

\newcommand{\de}{{\rm d}}
\newcommand{\vect}[1]{\boldsymbol{#1}}

\newcommand{\al}{\alpha}

\definecolor{grey}{rgb}{0.4,0.4,0.4}
\definecolor{dullmagenta}{rgb}{0.4,0,0.4}
\definecolor{darkblue}{rgb}{0,0,0.4}
\definecolor{midblue}{rgb}{0,0,0.5}
\definecolor{midred}{rgb}{0.5,0,0}
\definecolor{orange}{rgb}{1,0.5,0}
\definecolor{lightbrown}{rgb}{0.75,0.5,0.25}
\definecolor{tan}{cmyk}{0.14,0.42,0.56,0}
\definecolor{djunglegreen}{cmyk}{0.99,0,0.52,0}
\definecolor{lightgreen}{rgb}{0,1,0}
\definecolor{olivegreen}{cmyk}{0.64,0,0.95,0.40}
\definecolor{midgreen}{rgb}{0.0,0.675,0.0}
\definecolor{darkgreen}{rgb}{0,0.5,0}
\definecolor{ceruleanblue}{rgb}{0.0, 0.2, 0.7}
\definecolor{burgundy}{rgb}{0.5, 0.0, 0.13}

\hypersetup{
    colorlinks=true,
    linkcolor=ceruleanblue,
    filecolor=ceruleanblue,      
    urlcolor=midblue,
    citecolor=burgundy,
}

\newcommand{\mz}[1]{\textcolor{midgreen}{#1}}
\newcommand{\gt}[1]{\textcolor{midblue}{#1}}
\newcommand{\mzc}[1]{\textcolor{midgreen}{\textbf{[MZ: #1]}}}

\newcommand{\gtc}[1]{\textcolor{midblue}{\textbf{[GT: #1]}}}
\newcommand{\mcc}[1]{\textcolor{midred}{\textbf{[MC: #1]}}}
\newcommand{\mct}[1]{\textcolor{midred}{#1}}

\usepackage[all]{xy} 
\usepackage{amsfonts}

\begin{document} 

\title{Lensing of gravitational waves: efficient wave-optics methods and validation with symmetric lenses}

\author{Giovanni Tambalo}
\email{giovanni.tambalo@aei.mpg.de}
\affiliation{Max Planck Institute for Gravitational Physics (Albert Einstein Institute) \\
Am Mühlenberg 1, D-14476 Potsdam-Golm, Germany}

\author{Miguel Zumalac\'arregui}
\email{miguel.zumalacarregui@aei.mpg.de}
\affiliation{Max Planck Institute for Gravitational Physics (Albert Einstein Institute) \\
Am Mühlenberg 1, D-14476 Potsdam-Golm, Germany}

\author{Liang Dai}
\email{liangdai@berkeley.edu}
\affiliation{University of California at Berkeley,
Berkeley, California 94720, USA}

\author{Mark Ho-Yeuk Cheung}
\email{hcheung5@jhu.edu}
\affiliation{William H. Miller III Department of Physics and Astronomy, Johns Hopkins University, 3400 North Charles Street, Baltimore, Maryland, 21218, USA}

\begin{abstract}
Gravitational wave (GW) astronomy offers the potential to probe the wave-optics regime of gravitational lensing. Wave optics (WO) effects are relevant at low frequencies, when the wavelength is comparable to the characteristic lensing time delay multiplied by the speed of light, and are thus often negligible for electromagnetic signals.
Accurate predictions require computing the conditionally convergent diffraction integral, which involves highly oscillatory integrands and is numerically difficult.
We develop and implement several methods to compute lensing predictions in the WO regime valid for general gravitational lenses. First, we derive approximations for high and low frequencies, obtaining explicit expressions for several analytic lens models. 
Next, we discuss two numerical methods suitable in the intermediate frequency range: 1) \textit{Regularized contour flow} yields accurate answers in a fraction of a second for a broad range of frequencies. 2) \textit{Complex deformation} is slower, but requires no knowledge of solutions to the geometric lens equation. 
Both methods are independent and complement each other. We verify sub-percent accuracy for several lens models, which should be sufficient for applications to GW astronomy in the near future.
Apart from modelling lensed GWs, our method will also be applicable to the study of plasma lensing of radio waves and tests of gravity.
\end{abstract}

\date{\today}

\maketitle

{
  \hypersetup{hidelinks}
  \tableofcontents
}

\section{Introduction}

Gravitational lensing, the deflection of waves by gravitational fields, has become an essential tool for exploring the Universe's structure and contents. The rich phenomenology of gravitational lensing \cite{Bartelmann:2010fz} has enabled many applications, from inferring cosmological parameters to testing dark matter models. Progress on these fronts has relied exclusively on observations across the electromagnetic spectrum. However, recent advances in GW astronomy \cite{LIGOScientific:2016aoc,LIGOScientific:2021djp} open up a new arena for gravitational lensing. In time, searches of lensed GWs \cite{LIGOScientific:2021izm,Dai:2020tpj} are likely to turn into conclusive detections and novel applications. 

Differences between gravitational and electromagnetic radiation from astrophysical sources make GW lensing qualitatively distinct and complementary to electromagnetic observations.
GWs emit coherently and at much lower frequencies. 
GWs detectable by LIGO-Virgo-Kagra (LVK) have wavelengths more than three orders of magnitude longer than the lowest frequency radio waves permitted by the Earth's ionosphere.
This difference may allow the observation of WO effects \cite{Dai:2018enj, Oguri:2019fix}, which emerge when the wavelength is comparable to the time delay produced by the lens multiplied by the speed of light. WO effects are frequency dependent and their detection would allow an accurate determination of the lens properties \cite{Takahashi:2003ix,Cremonese:2021ahz,Caliskan:2022hbu,Savastano:2022jjv,Savastano:2023spl}. Moreover, WO lensing of GWs could serve to identify stellar-scale microlenses \cite{Christian:2018vsi,Diego:2019lcd,Cheung:2020okf,Mishra:2021xzz,Yeung:2021roe} and test dark matter scenarios \cite{Dai:2018enj,Jung:2017flg,Diego:2019rzc,Oguri:2020ldf,Basak:2021ten,Urrutia:2021qak,Guo:2022dre,Oguri:2022zpn}. 

Unfortunately, accurately computing lensed waveforms in the WO regime is challenging.
A closed-form analytical expression exists only for the simplest point-mass lens \cite{peters1974index}, and series expansions have been developed for more general lenses~\cite{Matsunaga:2006uc}.
These solutions have been used widely to study WO lensing.
However, even these simple expressions are costly to evaluate in practice, particularly at high frequencies. 
General predictions require conditionally convergent integrals of rapidly oscillating functions over the lens plane.
Previous works used direct integration \cite{Takahashi:2004mc,Dai:2018enj}, Levin's algorithm method \cite{Moylan:2007fi, Guo:2020eqw}, sampling the Fermat potential over contours \cite{Ulmer:1994ij,Mishra:2021xzz} (related to our first method) or discretely \cite{Diego:2019lcd, Cheung:2020okf, Yeung:2021roe}, discrete FFT-convolution \cite{Grillo:2018qjt}, and Picard-Lefschetz theory \cite{Feldbrugge:2019fjs,Jow:2022pux} (related to our second method). While these methods have been used to study complex lenses (e.g.~Refs \cite{Diego:2019lcd, Cheung:2020okf, Mishra:2021xzz, Yeung:2021roe}), they have been validated (e.g. cross-validated with independent calculations) only for simple examples.

Here we describe methods to obtain WO predictions and cross-validate them on several lens models. In Section \ref{sec:lensing_wo}, we present the diffraction integral. Section \ref{sec:lensing_expansions} presents expansions valid in the low- and high-frequency limits before turning to general, numerical algorithms.
In Section \ref{sec:wo_contour} we solve the Fourier transform of the integral by adaptively sampling contours of equal time delay. Then, in Section \ref{sec:wo_complex_def} we analytically continue the integration variable to make the integral manifestly convergent. Finally, in Section \ref{sec:wo_compare} we validate the accuracy of both methods and discuss their performance. We will explore the phenomenology of GW lensing separately \cite{our_paper}. Throughout this paper, we will work in a unit system with $c=1$.

\section{Wave Optics Regime of Gravitational Lensing} \label{sec:lensing_wo}

In this Section, we will review the equations governing gravitational lensing in the WO regime. 
In order to focus on the mathematical problem we will not provide a detailed derivation of the quantities involved. Readers are referred to Refs.~\cite{our_paper, Takahashi:2003ix} for details.
Our goal is to evaluate the diffraction integral, which we will give in dimensionless form:
\begin{equation}\label{eq:lensing_wave optics}
    F(w) 
    =
    \frac{w}{2\pi i}\int {\de}^2 \vect x\,e^{i w \phi(\vect x, \vect y)}
    \,.
\end{equation}
See Ref.~\cite{Schneider:1992} for a derivation. The integration is over the lens plane, with the coordinates rescaled by a dimensionful scale $\xi_0$ (e.g.~a characteristic scale of the lens), so $\vect x$ is dimensionless. The impact parameter $\vect y$ is rescaled by $\eta_0 \equiv D_S\xi_0/D_L$, where $D_S,\, D_L$ are the angular diameter distances to the lens and the source, respectively.

Here we introduced the \emph{dimensionless frequency}
\begin{equation} \label{eq:lensing_freq_dimensionless}
    w \equiv 8\pi G M_{Lz} f\,,
\end{equation}
which is given in terms of a redshifted \emph{effective lens mass}:
\begin{equation}
    M_{Lz} \equiv \frac{\xi_0^2 }{2 G d_{\rm eff}}\,.
\end{equation}
The factor $d_{\rm eff}\equiv \frac{D_L D_{LS}}{(1+z_L)D_S}$ also depends on the angular diameter distance between the 
lens
and the source $D_{LS}$. For a point lens, $M_{Lz}$ coincides with the total mass of the lens (i.e.~setting $\xi_0$ to be the Einstein radius), but this is not true for extended lenses.

The integral depends on the \emph{Fermat potential}: 
\begin{equation}\label{eq:fermat}
    \phi(\vect x, \vect y) 
    =
    \frac{1}{2}|\vect x - \vect y|^2- \psi(\vect x) - \phi_m(\vect y)\;.
\end{equation}
Here $\psi$ is the lensing potential, which depends on the matter distribution projected on the lens plane and whose derivative gives the deflection angle (Eq.~\eqref{eq:lens_eq} below). We conventionally shift by $\phi_m(\vect y)$, the global minimum value of the Fermat potential.
From here on, we will suppress $\phi_m(\vect y)$ in our formulas and assume that it is added to make the minimum arrival time equal to zero. When necessary, we will introduce it back.

\begin{figure}
    \centering
    \includegraphics[width=\columnwidth]{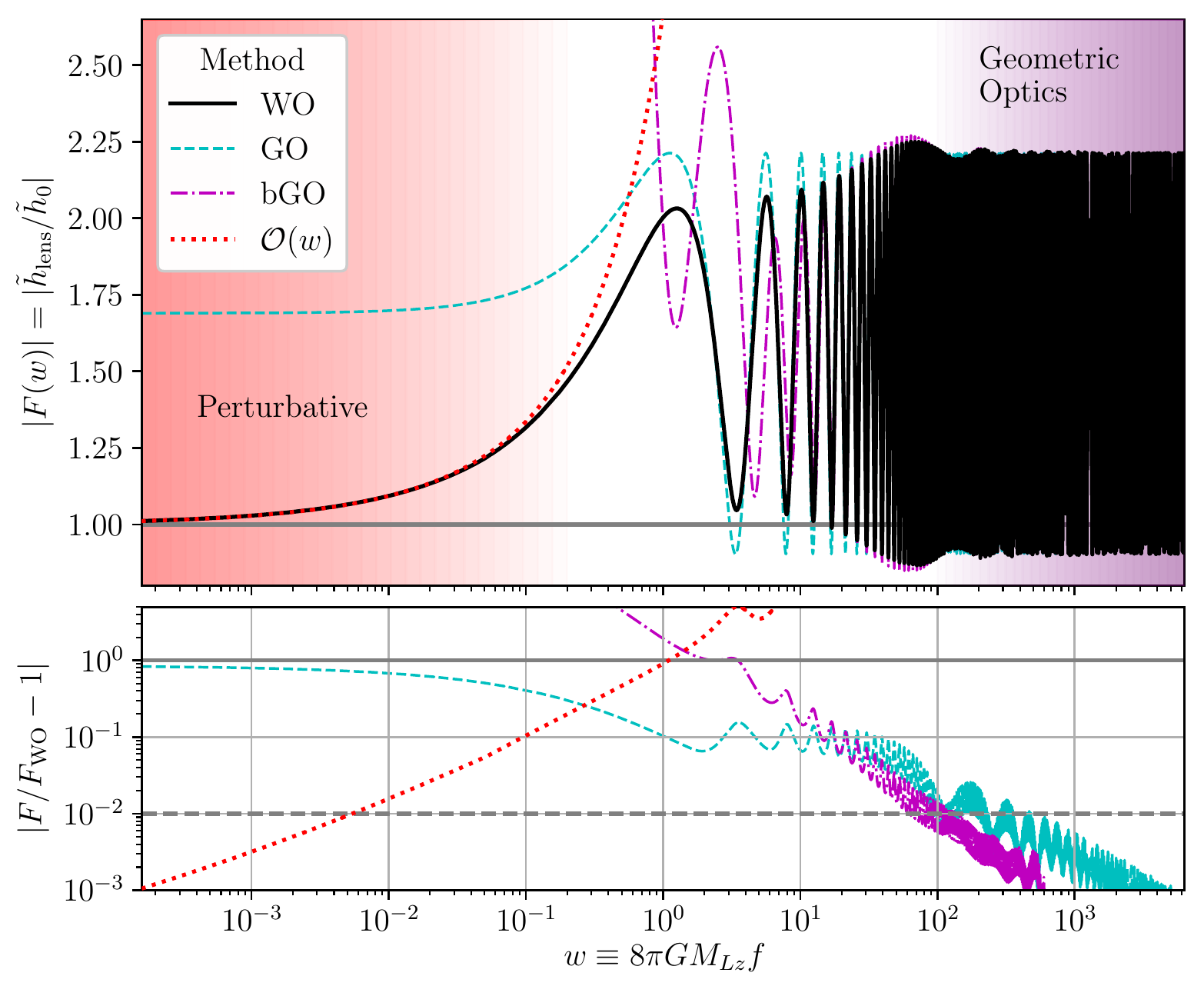}
    \caption{Amplification factor for an SIS lens ($y=0.3)$. \textbf{Top panel:} The full WO solution (solid) is compared to different approximations: low $w$ series expansion (dotted), geometric optics (dashed) and beyond geometric optics (dash-dotted). The regions where systematic expansions are good descriptions appear shaded.
    {\textbf{Bottom panel:} Relative deviation with respect to the WO solution.}
    \label{fig:lensing_regions}}
\end{figure}

We will consider several lens models in this work, summarized in Table \ref{tab:lenses_summary}. 
All of them are spherically symmetric, leading to an axi-symmetric projected mass and lensing potential $\psi(\vect x) = \psi(x)$ (here and in the following $x \equiv |\vect x|$).
First, we consider the point lens, whose analytic solution will help us test the accuracy of different methods in Sec.~\ref{sec:wo_compare}. 
We will additionally consider three extended lenses: the Singular Isothermal Sphere (SIS), and two one-parameter extensions. SIS lenses follow from a matter profile $\rho\propto 1/r^2$ and are often employed to model lensing by galaxies. 
Our first extension, the generalized-SIS (gSIS), has an arbitrary slope $\rho\propto 1/r^{1+k}$ ($0<k<2$) and can be used to model steeper or shallower lenses \cite{Schneider:1992, Keeton:2001ss, Choi:2021jqn}. Its central density diverges, but the enclosed mass up to some radius remains finite if $k<2$.
Our second extension, the Cored Isothermal Sphere (CIS), has a central core of physical radius $r_c=x_c \xi_0$ \cite{1987ApJ...320..468H,Flores:1995dc}. Therefore, the matter density $\rho \propto 1/(r^2+r_c^2)$ is finite at the centre. Details about these lenses and their phenomenology will be provided separately \cite{our_paper}.

\begin{table*}[t]
    \centering
    \setlength{\tabcolsep}{10pt}
        \begin{tabular}{l c  c  c }
         Name & $\rho(r)$ & $\psi(x)$ & Parameters 
         \\[0.2cm] \hline 
         Point Lens & $\delta_D(r)$ & $\log(x)$ & - 
         \\ [0.15cm]
         Singular Isothermal Sphere (SIS) & $\frac{1}{r^{2}}$ & $x$ & - 
         \\[0.15cm]
         Generalized SIS (gSIS) & $\frac{1}{r^{(k+1)}}$ & $\frac{x^{(2-k)}}{(2-k)}$ 
         & Slope $k$ 
         \\[0.15cm]
         Cored Isothermal Sphere (CIS) & $\frac{1}{r^2+r_c^2}$ & 
         $\sqrt{x_c^2+x^2}+x_c \log \left(\frac{2 x_c}{\sqrt{x_c^2+x^2}+x_c}\right)$ 
         & Core size $x_c$ 
         \\[0.3cm] \hline
    \end{tabular}
    \setlength{\tabcolsep}{6pt}
    \caption{Summary of lens models used in this work. For reference, we have included $\rho(r)$, the density of a spherically symmetric matter distribution leading to the lensing potential $\psi$ in each case.
    }
    \label{tab:lenses_summary}
\end{table*}

Figure \ref{fig:lensing_regions} shows the amplification factor for an SIS lens with impact parameter $y=0.3$. The full WO solution was obtained by regularized contour flow, discussed in Sec.~\ref{sec:wo_contour}, matched to an analytic expansion at high frequencies. The remaining curves correspond to the expansions presented in Sec.~\ref{sec:lensing_expansions}, each with limited range of validity: Geometric Optics (GO) and its next-order refinement (bGO) (Sec.~\ref{sec:lensing_go}) are good descriptions at high frequency, while the series expansion (Sec.~\ref{subsec:wo_low}) is a good approximation only for $w\ll 1$. GO remains bounded at all frequencies, while bGO and the series expansion diverge at low/high frequencies, respectively.

\section{Analytic Expansions} \label{sec:lensing_expansions}

We now present analytic expansions valid in the high- and low-frequency limits, Subsections \ref{sec:lensing_go}, \ref{subsec:wo_low}, respectively. 

\subsection{Geometric optics \& beyond}
\label{sec:lensing_go}
In the high-frequency limit, following the same arguments leading to the stationary-phase approximation for path integrals, only the neighbourhoods of extrema of the Fermat potential \eqref{eq:fermat} contribute to the amplification factor \eqref{eq:lensing_wave optics}. 
Each extremum is associated with an image $J$, located at a position $\vect x_J$ in the image plane where the lens equation
\begin{equation}\label{eq:lens_eq}
    \vect \nabla_{\vect x} \phi(\vect x_J, \vect y) 
    =
    \vect x_J - \vect y - \vect\alpha(\vect x_J) 
    =
    0
    \,,
\end{equation}
is satisfied (here $\vect \alpha(\vect x_{J})\equiv \vect \nabla_{\vect x} \psi(\vect x_J)$ is the deflection angle and $\vect \nabla_{\vect x} $ is the gradient computed with respect to $\vect x$).
The \textit{geometric optics} regime emerges from a quadratic expansion of the Fermat potential around each image so that the diffraction integral can be performed analytically. 

The GO amplification factor \eqref{eq:lensing_wave optics} receives contributions from each image $J$
\begin{equation}\label{eq:lensing_geometric_optics}
    F(w) 
    =
    \sum_J |\mu_J|^{1/2}\,e^{i w \phi_J}\,e^{- i\pi n_J}
    \,,
\end{equation}
where the \emph{magnification}
\begin{equation}\label{eq:go_magnification_def}
    \mu^{-1} 
    \equiv 
    \det\left(\phi_{,ij}\right)
    =
    \left(1-\frac{\alpha(x)}{x}\right)\left(1-\frac{\de \alpha(x)}{\de x}\right)
    \,,
\end{equation}
is evaluated on the image position $x_J$ (the second equality above applies to the specific case of axially-symmetric lenses).
In the above expressions, $\phi_J$ is the Fermat potential of the $J$-th image, $\phi_{, ij} \equiv \partial_i\partial_j \phi$ is its Hessian matrix and $\alpha(x) \equiv |\vect \alpha(\vect x)|$. As we are working in the two-dimensional lens plane, $i, j, \dots \in \{1, 2 \}$, corresponding to the $x_1$ and $x_2$ coordinates (the Cartesian components of $\vect x$). 
The \textit{Morse Phase} \cite{Schneider:1992,Takahashi:2003ix, Dai:2017huk} depends on the type of image as
\begin{equation}
    n_J = \left\{
    \begin{array}{lll}
  0 & \;\text{if} \det\left(\phi_{,ij}\right), \tr\left(\phi_{,ij}\right)>0 & \text{(minima)} \\[3pt]
  \frac{1}{2}& \;\text{if} \det\left(\phi_{,ij}\right) <0 & \text{(saddle)} \\[3pt]
  1 & \;\text{if} \det\left(\phi_{,ij}\right) >0\,,\, \tr\left(\phi_{,ij}\right)<0 & \text{(maxima)} 
    \end{array}
    \right.\,.
\end{equation}
Minima, saddle points and maxima of the time delay function are also known as type I, II and III images, respectively \cite{Blandford:1986zz}.

\subsubsection{Beyond geometric optics}

Beyond GO (bGO) corrections can be obtained as a series expansion in $1/w$.  We now review the leading bGO correction, following \cite{Takahashi:2004mc} and focusing on axially-symmetric lenses (see also \cite{takahashiThesis}). 
First of all, we expand the lensing potential $\phi$ around each image $\vect x_{J}$ up to quartic order in $\tilde x^i \equiv x^i-x^i_{J}$:
\begin{align}
    \phi(\vect x, \vect y)
    &
    =
    \phi_J 
    + \frac{1}{2} (\phi_J)_{,ij} \, \tilde x^i \tilde x^j 
    + \frac{1}{3!}(\phi_J)_{,ijk} \, \tilde x^i \tilde x^j \tilde x^k
    \nn \\  
    &
    + \frac{1}{4!}(\phi_J)_{,ijkl} \, \tilde x^i \tilde x^j \tilde x^k \tilde x^l
    + \mathcal O (\tilde x^5)\;. 
    \label{eq:taylor_lens_pot}
\end{align}

For a symmetric lens, the quadratic term in $\tilde x^i$ is diagonal and can be written as (here $\vect y$ is taken to be along the $x_1$ direction so that $x^i_J = x_J \delta_1^i$)
\begin{align}
    (\phi_J)_{,ij} \, \tilde x^i \tilde x^j 
    &
    =
    (1-\psi_J'')\tilde x_1^2 
    + \left( 1-\frac{\psi_J'}{x_J}\right) \tilde x_2^2 
    \nn \\
    & = 
    2 a_J\tilde x_1^2 
    + 2 b_J \tilde x_2^2 \;,
\end{align}
where primes denote radial derivatives and we defined $a_J \equiv (1-\psi_J'')/2$ and $b_J \equiv ( 1-\psi_J'/x_J)/2$.
At this point, in the diffraction integral Eq.~\eqref{eq:lensing_wave optics} we shift and rescale $x^i$ to $z^i \equiv \sqrt{w} \tilde x^i$. 
In terms of $z^i$, the quadratic term at the exponent is $w$ independent. The cubic and quartic terms of Eq.~\eqref{eq:taylor_lens_pot}, multiplied by $w$, scale instead as $\mathcal O(w^{-1/2})$ and $\mathcal O (w^{-1})$ respectively. For large $w$, we can then Taylor expand the exponential and keep terms up to $\mathcal O(w^{-1})$.

At order $\mathcal O(w^0)$ we recover the GO result (from the quadratic part). The term of order $\mathcal O(w^{-1/2})$ vanishes since it leads to an odd integrand. The first correction comes instead at order $\mathcal O (w^{-1})$, where we have two distinct contributions: one from squaring the term $(\phi_J)_{,ijk} \, \tilde x^i \tilde x^j \tilde x^k$ and another from $(\phi_J)_{,ijkl} \, \tilde x^i \tilde x^j \tilde x^k \tilde x^l$. 
Terms with higher powers of $\tilde x$ contribute at order $\mathcal O (w^{-3/2})$ or higher, and can thus be neglected at sufficiently large frequencies.

After performing these two integrals, one is left with the following simple result
\begin{equation}
    F(w) 
    = 
    \sum_{J} |\mu_J|^{1/2} 
    \left(1+ i\frac{\Delta_J}{w}\right)
    e^{i w \phi_J - i \pi n_J} 
    + \mathcal O(1/w^2)\;,
\label{eq:bGO}
\end{equation}
where the real number $\Delta_{J}$ characterizes the bGO correction, and is given by
\begin{equation}
    \Delta_{J}
    \equiv
    \frac{1}{16}\left[
    \frac{\psi_{J}^{(4)}}{2 a_{J}^{2}}
    +\frac{5}{12 a_{J}^{3}} (\psi_{J}^{(3)})^{2}
    +\frac{\psi_{J}^{(3)}}{a_{J}^{2} x_{J}}
    +\frac{a_{J}-b_{J}}{a_{J} b_{J} x_{J}^{2}}
    \right]\;.
\label{eq:Delta_bGO}
\end{equation}
Here $\psi^{(n)}\equiv \frac{\de^n}{\de x^n}\psi$. 

Equation \eqref{eq:bGO} shows that the leading-order GO result is a good approximation provided that $\Delta_J/w\ll1$ for all images.\footnote{Another GO convergence criterion is that $w(\phi_I-\phi_J)\gg 1\,,\forall I\neq J$. This can be understood from the contours framework (\ref{sec:wo_contour}) as the images being resolvable at finite frequency.
Note that this criterion is, in general, independent from bGO terms being negligible, $\Delta / w \ll 1$.
} Note also that non-analytic features in the Fermat potential (e.g.~cusps) produce other $\mathcal{O}(w^{-1})$ contributions without a corresponding GO image \cite{Takahashi:2004mc}. 
We will now address the contribution of non-analytic features in specific cases.

\subsubsection{Contribution from the cusp}\label{subsubsec:cusp_features}

The leading terms in the GO expansion, Eq.~\eqref{eq:lensing_geometric_optics}, arise from the stationary points of the Fermat potential and capture the high-$w$ contributions to the amplification factor. Nonetheless, other locations in the lens plane can induce corrections at subleading order in the $\sim 1/w^n$ expansion and might be comparable to the bGO terms.
In particular, they can arise from singular points of the lens equation (cusps in the lensing potential). See \cite{Takahashi:2004mc} for a similar discussion on cusp contributions to $F(w)$.

In this Subsection we are going to discuss these contributions for the lens models featuring a central cusp (gSIS and SIS lenses, Table \ref{tab:lenses_summary}).\footnote{The centre is smooth for the CIS lens, so no new contribution arises compared with bGO. The point lens is singular at the centre, but the new contribution is highly suppressed in $w$ (see \cite{Takahashi:2004mc} for a discussion). Thus, the gSIS is the only relevant case among the lenses we consider.} In particular, we focus on the strong lensing regime, where $y$ can be taken as a small number.

Let us consider the gSIS lens in the limit of large $w$. For this lens, we distinguish two behaviours depending on the value of the slope $k$. For $0<k<1$ (broad profiles) the lens equation is smooth at the lens' centre and a central image forms in the strong-lensing regime. 
In other terms, the deflection angle $\alpha$ is bounded as the ray approaches the centre of the lens.
In the complementary interval $1\leq k<2 $ (narrow profiles) the lens equation is singular at the centre and no image forms.
In both these cases we isolate the contribution to $F(w)$ from the centre by truncating the integration range from $x\in(0, R_c)$, for some radius $R_c$ small enough for the GO images not to be enclosed. The range $x>R_c$, at high $w$, is then dominated by the GO expansion around the minimum and/or saddle (depending on $y$). In the lower integration interval, we have instead
\begin{align}
    F_c(w)
    &=
    \frac{w}{2\pi i }\int_0^{R_c}\de x \, x \int_0^\pi \de \theta \,
    e^{i w \phi(\vect x, \vect y)}
    \nn \\
    &=
    -i w e^{i w\phi_c}\int_0^{R_c} \de x \, x J_0(wyx) e^{i w (\frac{x^2}{2}- \psi(x))}\;,
  \label{eq:F_cusp}
\end{align}
where $\phi_c \equiv y^2/2 - \phi_m$ is the time delay associated to the lens centre (here we re-introduced the minimum time delay $\phi_m$) and $J_\nu(z)$ is the Bessel function of the first kind (obtained after performing the angular integral). The integrand, in the limit $w\gg1$, peaks around $x = 0$ once we rotate the integration line into the complex plane. To see this, first notice that $J_0(wyx) e^{i w x^2/2} \simeq 1$ for small $x$ (we will motivate better why $x$ can be taken small \emph{a-posteriori}). By writing $x\, e^{-iw\psi(x)} = e^{\log x - iw \psi(x)} \equiv e^{\Omega}$, we can locate the peak as the stationary point $x_s$ of $\Omega$:
\begin{equation}
\label{eq:cusp_saddle_eq}
    \frac{\de }{\de x}\Omega 
    =
    \frac{1}{x_s} - i w x_s^{1/A-1} = 0
    \;.
\end{equation}
Here we defined $A \equiv 1/(2-k)$, which is a positive quantity.
Equation \eqref{eq:cusp_saddle_eq} is solved for $x_s = (i w)^{-A}$. Notice that $x_s$ becomes smaller for larger $w$, making our approximation adequate in this limit (in particular the Gaussian part at the exponent can be neglected since $x_s^2\ll \psi(x_s)$). 

Therefore, $F_c(w)$ for $w\gg1$ can be obtained using a saddle-point approximation around $x_s$. However, we prefer to take a slightly different approach that yields very similar results: we evaluate $J_0(wyx)e^{iw x^2/2}$ in Eq.~\eqref{eq:F_cusp} at the peak $x_s$, while performing the exact integration over for $e^{\Omega}$. Since the integral is highly localized for $w\gg1$, the calculation can be simplified by taking $R_c\rightarrow\infty$, making exponentially small errors.
We obtain
\begin{align}
    F_c(w)
    &\simeq 
    -i w e^{i w\phi_c}J_0(wyx_s)e^{iw x_s^2/2} \int_0^{\infty}  \de x \, x \,e^{-i w \psi(x)}
    \nn \\
    &=
    -e^{i w\phi_c}(i w A)^{1-2A} \Gamma(2A) J_0(wyx_s) e^{iw x_s^2/2}\;.
  \label{eq:F_c_sol}
\end{align}
This formula is valid when the limits $w\gg1$ and $w yx_s\ll1$ are satisfied (therefore, for small enough $y$).%
\footnote{Contribution from cusps are computed in \cite{Takahashi:2004mc} for SIS and gSIS lenses. For the SIS, Eq.~\eqref{eq:F_c_sol} matches with 
Eq.~(21) of \cite{Takahashi:2004mc} for small $y$. For the generic gSIS instead, 
the reference implicitly assumes large $y$, so that our formulas cannot be directly compared.
}
The full $F(w)$ is then given by the sum of Eq.~\eqref{eq:F_c_sol} and the usual GO expansion for the other images. In the following, we will refer to this expansion as \emph{resummed} GO (rGO).
We can better understand the behaviour of $F_c(w)$ by first looking at the SIS case $k=1$. Here, neglecting again the Gaussian and the Bessel function, we have $F_c(w) \simeq i/w e^{i w \phi_c}$. 
This can be interpreted as an additional bGO contribution from the cusp $x = x_c = 0$, with time delay $\phi_c$ and with a vanishing GO term (i.e.~not accompanied by an image).
More in general, for narrow (broad) profiles, $F_c(w)$ decays faster (slower) than $1/w$.

For broad profiles, there is a caveat in the previous derivation at very large $w$: the Gaussian part can start contributing significantly to the integral, thus leading to the usual GO expansion for the central image. Therefore, for $0<k<1$, $F_c$ in Eq.~\eqref{eq:F_c_sol} is a better approximation than bGO for the central image only in the range $1\lesssim w \lesssim \Delta_c$, while for $\Delta_c/w\ll1$ bGO performs better (here $\Delta_c$ is the bGO coefficient of the central image, Eq.~\eqref{eq:Delta_bGO}). This issue does not arise for $k>1$, since here there is no central image.

From the discussion above we conclude that WO effects from the cusp are relevant even when no central image forms. As we will study in \cite{our_paper}, this has interesting implications for parameter estimations with GWs.

\subsection{Low-frequency expansion} \label{subsec:wo_low}

We are now interested in understanding the behaviour of the amplification factor in the limit of small $w$. In this limit, GO fails and one has to resort to other methods to obtain good approximations.

For small $w$, $F(w)$ approaches $1$ since the wavelength becomes much larger than the lens' characteristic scale, and the wave is unperturbed by the lens.
Here we would like to motivate that corrections to $F(w) \sim 1$ in this limit correspond to an expansion in powers of the lensing potential $\psi(\vect x)$.
A physical motivation can be given as follows.
If the wavelength is much larger than the typical scale of the lens (i.e.~Einstein radius), then 
the impact parameter's value cannot be precisely resolved. This implies that the impact parameter should be irrelevant (at least at leading order) in this low-frequency limit. Thus, we could imagine performing the calculation for $F(w)$ with $y\gg1$ (i.e.~impact parameter much larger than the scale of the lens, set by $\xi_0$) but still much smaller than the wavelength $\propto 1/w$. In this case one can expand the diffraction integral in powers of the lensing potential (see \cite{Takahashi:2005sxa} on the conditions for the applicability of this approximation).

We can also see explicitly that this procedure gives a sensible series expansion in $w$: higher powers on $\psi(\vect x)$ lead to subleading terms in $w$. For simplicity, we show this for axis-symmetric lenses.

First, we perform a rotation of the integration contour to make the integrals manifestly convergent.
Similarly to Eq.~\eqref{eq:F_cusp}, the amplification factor can be written as follows
\begin{align}\label{eq:gaussian_contour}
    & F(w) 
    =
    -i w
    e^{i w y^2/2}
    \int_0^{\infty} \de x \, 
    x
    J_0(w y x)
    e^{i w (x^2 / 2 - \psi(x))} 
    \nn \\
    &= 
    e^{i w y^2/2}
    \int_0^{\infty} \de z \,
    z
    J_0(e^{i \pi / 4} \sqrt{w} y z)
    e^{-z^2 / 2 - i w \psi(x(z)))} 
    \;,
\end{align}
where in the second line we rescaled the radial variable and rotated the integration contour by 45 degrees in the complex plane: $x = e^{i \pi /4} z / \sqrt{w}$.\footnote{Here we implicitly assumed $\psi(x)$ to be analytic in the region $0 \leq {\rm arg}\, x < \pi/2$ of the complex-$x$ plane. For the lenses we consider in Tab.~\ref{tab:lenses_summary} this is the case and it is possible to perform the 45-degrees rotation in the complex plane without hitting singularities. For more general situations Eq.~\eqref{eq:step1_low_w} needs to be modified to include the contribution of singularities.}
Note that the Gaussian part dictates the leading behaviour at infinity of the integrand, since the Bessel function only grows exponentially and we assume for convergence that $|\psi|$ grows more slowly than the argument of the Gaussian ($\lim_{x\to\infty}|\psi|/x^2=0$, see Eq.~\eqref{eq:low_w_convergence} below). Thus, convergence is now manifest.
However, this choice for the integration contours is not optimal: the Bessel function and the lensing potential can make the integrand exponentially large at intermediate values of $z$, for $w$ large.
On the other hand, for small $w$, Eq.~\eqref{eq:gaussian_contour} is easy to evaluate, giving us a tool to explore $F(w)$ in the WO regime. 
We will discuss how to obtain the optimal integration contour for more general lenses in Sec.~\ref{sec:wo_complex_def}.

After this first step, we can expand the integrand in powers of  $\psi(x(z))$.
Equation \eqref{eq:gaussian_contour}, expanded in $\psi(x(z))$ up to quadratic order, has the form
\begin{align}\label{eq:step1_low_w}
    F(w) 
    & \simeq
    1 
    - e^{i w y^2/2}
    \int_0^\infty \de z \, 
    z \, e^{-z^2/2} J_0(e^{i \pi /4} \sqrt{w} y z)  
    \nonumber \\
    & \times
    \left[
    i w  \psi\left( \frac{e^{i \pi /4} z}{\sqrt{w}}\right)  
    + \frac{w^{2}}{2} \psi\left( \frac{e^{i \pi /4} z}{\sqrt{w}}\right)^2 
    \right]
    \;.
\end{align}
Obtaining higher terms in this expansion is also trivial. 
If $\psi$ is well behaved for large $z$, then the integrals in Eq.~\eqref{eq:step1_low_w} are localized around $z = 1$ and can be estimated through a steepest-descent calculation.\footnote{Note that this is just an approximation we can use to estimate the integrals, and does not reduce to the exact answer in any limit. 
This approximation is often not required for specific lens models, as one can just evaluate the integrals exactly.} See \cite{Choi:2021jqn} for a similar expansion, but in the weak-lensing regime.

Let us first consider $y = 0$ for simplicity. With the $z=1$ approximation we obtain
\begin{equation}
    F 
    \simeq 
    1 - i w \psi\left(\frac{e^{i\pi/4}}{\sqrt{w}}\right) 
    - \frac{w^2}{2}\psi\left(\frac{e^{i\pi/4}}{\sqrt{w}}\right)^2
    \;.
\end{equation}
This is a series in power of $w \psi\left(e^{i\pi/4} / \sqrt{w}\right)$: higher powers in the expansion are suppressed for small $w$ provided that 
\begin{equation}\label{eq:low_w_convergence}
    \lim_{w\rightarrow 0} w \psi\left(e^{i\pi/4} / \sqrt{w}\right) = 0\,.    
\end{equation}
This condition is physically sensible since it is equivalent to the requirement for $\psi(x)$ to grow less than the quadratic part of the lensing potential at infinity (in this particular direction of the complex plane). Equivalently, in terms of the density profile $\rho$, the requirement translates to $\rho$ decaying faster that $\propto 1/r$ at infinity.
All the analytic lens models we consider in this work satisfy this requirement, hence this expansion is applicable in our cases. 

The conclusion that Eq.~\eqref{eq:step1_low_w} is a good expansion for small $w$ is not spoiled in the case of $y \neq 0$. To see this, we first notice that $y$ enters only the combination $w y^2$ in Eq.~\eqref{eq:step1_low_w}. Hence, if $w y^2$ is smaller than $1$, we can expand the Bessel function in a series around zero (the integral is localized around $z = 1$ and the argument of $J_0$ remains small). In the limit $w \ll 1$ this is possible, with the mild requirement of $y \ll 1/\sqrt{w}$. Additionally, we can also notice that $y$ will not enter at first order in the $w$ expansion of $F(w)$. Indeed, the series expansion $J_0(e^{i \pi /4} \sqrt{w} y z) \simeq 1 -\frac{i}{4} w y^2 z^2 + \mathcal O (w^2 y^4 z^4)$ shows that effects due to $y$ are suppressed by additional powers of $w$.

\subsubsection{Leading corrections for symmetric lenses} \label{subsubsec:wo_low}

After these general results, we can focus on the low-$w$ behaviour for the specific lenses shown in Table \ref{tab:lenses_summary}. For the simplest cases, we can directly integrate Eq.~\eqref{eq:step1_low_w} without assuming small $y$.

\textbf{Point lens}: 
In this case $\psi(x) = \log x$ and the integral in Eq.~\eqref{eq:lensing_wave optics} has a closed-form solution \cite{Takahashi:2003ix}:
\begin{align}\label{eq:pt_lens_analytic}
    F(w) 
    &= 
    e^{\frac{\pi}{4}w+ i \frac{w}{2}(\log\left(\frac{w}{2}\right)-2\phi_m)} \Gamma\left(1-i \frac{w}{2}\right) 
    \nn \\
    & 
    \times {}_1F_1 \left(\frac{iw}{2},1, i w \frac{y^2}{2}\right)\;,
\end{align}
where $\phi_m = (x_m-y)^2/2-\log(x_m)$ is the minimum of the Fermat potential, evaluated at $x_m = (y+\sqrt{y^2+4})/2$. Moreover, ${}_{1} F_{1}(a, b, z)$ is the confluent hypergeometric function. 
We can nonetheless employ the low-frequency expansion and then compare with the formula above, expanded in the same limit. 
Equation \eqref{eq:step1_low_w} (neglecting $w^2 \psi^2$ terms) gives
\begin{equation}
    F^{\rm pl} 
    \simeq 
    1 
    +\frac{w}{2}
    \left[
    \pi - i (2\log y - {\rm{Ei}}(i w y^2/2))
    \right] 
    + \mathcal O (w^2)
    \;,
\label{eq:pl_low_w}
\end{equation}
where ${\rm Ei}(z)$ is the exponential integral. 
We checked that indeed Eq.~\eqref{eq:pt_lens_analytic} expanded for small $w$, but fixed $wy^2$, yields Eq.~\eqref{eq:pl_low_w}.
It is also useful to further expand Eq.~\eqref{eq:pl_low_w} for small $w y^2$:
\begin{equation}\label{eq:pl_low_w_low_y}
    F^{\rm pl} 
    \simeq 
    1 + \frac{w}{4}
    \left( 
    \pi + 2 i \gamma_{\rm E} + 2 i \log \frac{w}{2} 
    \right) 
    + \mathcal O (w^2  y^2)
    \;,
\end{equation}
where $\gamma_{\rm E}$ is the Euler's constant.

Let us briefly digress and comment about the analytic properties of Eq.~\eqref{eq:pl_low_w_low_y} as a function of $w$.
We can first note that the diffraction integral, Eq.~\eqref{eq:lensing_wave optics}, is analytic for ${\rm Im}\, w >0$, while possible non-analyticities can appear on the real axis and in the lower half of the complex plane.
The analyticity property for ${\rm Im}\, w >0$ is a consequence of causality: while the lensed signal can have a distorted waveform due to diffraction, no signal can travel beyond the light cone (see \cite{Suyama:2020lbf,Ezquiaga:2020spg} for a related discussion).
The above properties can be checked for the point lens, using its closed-form solution Eq.~\eqref{eq:pt_lens_analytic}. The latter formula is indeed analytic for ${\rm Im}\, w >0$, has poles due to the Gamma function for ${\rm Im}\, w < 0$ and a branch cut due to the $\log w$ for negative $w$.
Notice that by construction $F(w)$ satisfies a reality condition $F^*(w) = F(-w)$. Non-analyticity of $F(w)$ and the reality condition imply that $F(w)$ must be a complex number for $w$ real. Otherwise it would be either real or imaginary. 
By direct inspection, Eq.~\eqref{eq:pl_low_w_low_y} is consistent with these properties. In particular, a branch-cut 
appears due to $\log w$. Similar considerations apply to the lens models of Tab.~\ref{tab:lenses_summary}.

\textbf{SIS lens}: 
Here $\psi(x) = x$ and keeping up to the $\psi^2$ term in Eq.~\eqref{eq:step1_low_w} we obtain (again keeping $wy^2$ fixed)
\begin{align}
    F^{\rm SIS}  
    & \simeq
    1 
    - \frac{e^{i \pi/4}}{4} \sqrt{2 \pi w} e^{i wy^2/4} 
    \bigg[ 
    2i \left(1-\frac{i wy^2}{2}\right)  J_0(wy^2/4) 
     \nonumber \\
    & - i w y^2 J_1(w y^2/4) 
    \bigg]
    -\frac{w}{2}(2i + wy^2)+ \mathcal O (w^{3/2})\;.
\end{align}
After expanding the expression above for small $y$ we obtain
\begin{equation}
    F^{\rm SIS} 
    \simeq 
    1-(-1)^{3/4} \sqrt{\frac{\pi w}{2}} -i w 
    + \mathcal O (w^{3/2})
    \;.
\label{eq:low-w_SIS_low-y}
\end{equation}
Interestingly, in this case the leading behaviour in $w$ differs from the one of the point-lens. 
The dependence on $\sqrt{w}$ can be understood from Eq.~\eqref{eq:step1_low_w}. If we call $s \equiv \sqrt{w} y$ and use $\psi(x) = x$, we see that the only dependence on $w$ is through $\sqrt{w}$. Together with the fact that $s$ will not appear at leading order, we obtain the correct dependence of $F$.
The slope as a function of $\sqrt{w}$ is related to the steepness of $\psi(x)$ far from the centre of the lens. This can be seen by considering a generalized SIS lens, with a generic slope.

\textbf{Generalized SIS lens}: 
The gSIS lens, as already discussed, is described by the lensing potential $\psi(x) = x^{2-k}/(2-k)$, where the slope is parametrized by $0<k<2$. The case $k = 1$ reduces to the SIS.

The first-order expansion in $w$ leads to 
\begin{align}\label{eq:low-w_gSIS}
  F^{\rm gSIS} 
  &\simeq 
  1 + \left(-w/2\right)^{k/2}  \Gamma\left(1/2A\right) e^{i w \frac{y^2}{2}}L_{1/2A}(i w y^2/2) \nonumber \\
  & + \mathcal O (w)\;,
\end{align} 
where $\Gamma(z)$ is the Gamma function, $L_{\alpha}(z)$ are the Laguerre polynomials and we introduced again $A = 1/(2-k)$.
Expanding for small $y$ leads simply to  
\begin{equation}\label{eq:low-w_gSIS_low-y}
    F^{\rm gSIS} 
    \simeq 
    1 + \left(-w/2\right)^{k/2}  \Gamma\left(1/2A\right)
    + \mathcal O (w)
    \;.
\end{equation} 
This expression agrees with what is found in \cite{Choi:2021jqn}. Notice that the leading term in $w$ depends on the slope $k$: steep lenses ($k$ close to zero) have a weaker dependence on $w$.

\textbf{CIS lens}: 
This lens represents a SIS lens with a central core of size $x_c$. Its lensing potential is given in Table \ref{tab:lenses_summary}. 
As we have argued, in the limit of low $w$ we are mainly sensitive to the profile of the lens far away from the centre. Thus, we expect to have a weak dependence on $x_c$ in this limit. We do not know a closed expression for the integrals of $\psi(x)$ in Eq.~\eqref{eq:step1_low_w}. However, we can approximate them by noticing that since the integral is peaked around $z \sim 1$ and we are interested in $w \ll 1$, we can simply expand $\psi\left(e^{i \pi/4} z / \sqrt{w}\right)$ for large argument. Interestingly, for large $x$ we have $\psi(x)\simeq x-x_c \log \frac{x}{2x_c} + \mathcal O (1/x)$. In other words, the lens looks like an SIS plus a central point lens with negative mass. Notice that to correctly capture the dependence on $x_c$ we need to include the second-order correction from Eq.~\eqref{eq:step1_low_w}. By doing so, we obtain
\begin{align}
\label{eq:low_w_CIS_full}
    F^{\rm CIS} 
    \simeq &
    1 
    +(F^{\rm SIS}-1)- x_c (F^{\rm pl}-1) -
    \nonumber \\ 
    & - i w e^{-i w y^2/2} 
    \left[ 
    x_c \log(2 x
    _c) +1-i w y^2/2
    \right] 
    \;.
\end{align}
More explicitly, further expanding for small $y$ we have
\begin{align}
\label{eq:low_w_CIS_simpl}
    F^{\rm CIS} 
    &\simeq
    1 
    -\left[ 
    (-1)^{3/4} \sqrt{\frac{\pi w}{2}} 
    + \frac{w x_c}{4}\left( 2 i \gamma_{\rm E} 
    \right. \right. 
    \nn \\
    &\left. \left.
    + \pi 
    + 2 i  \log \frac{w}{2} +4 i \log (2x_c) + \frac{4 i }{x_c} 
    \right) \right]
    \;.
\end{align}
This result shows that the leading behaviour in $w$ is due to the $\rm SIS$ part, while the presence of the core is only visible at subleading order.
From Eq.~\eqref{eq:low_w_CIS_simpl}, we notice that increasing $x_c$ leads to a \emph{smaller} $|F(w)|$. This is consistent with our numerical results.
Later, we will test the accuracy of this approximation against the full WO result (Sec.~\ref{sec:compare_methods}, Fig.~\ref{fig:low_w_comparison_cis}).

Throughout this discussion we suppressed the minimum time delay $\phi_m$ (see discussion below Eq.~\eqref{eq:fermat}). To reintroduce this parameter, all the expanded expressions for $F(w)$ in this Section should be multiplied by the phase $e^{-i w \phi_m}$.

\subsubsection{gSIS series expansion}\label{subsubsec:series_expansion}

The amplification factor for simple lens models can be written in a useful series representation, by expanding the integrand in powers of the lensing potential $\psi(x)$. In particular, we will be able to obtain such representation for the SIS and gSIS lens models (see Tab.~\ref{tab:lenses_summary}). This provides an additional independent test of our numerical methods, which will be very valuable to validate our results.
Let us first consider a generic axially-symmetric $\psi(x)$ and later specialize to particular functional forms.
To proceed, we start from Eq.~\eqref{eq:lensing_wave optics} and perform the angular integral, which yields the usual Bessel function $J_0(wyx)$ (already encountered in Eq.~\eqref{eq:step1_low_w}). 
After this, we expand $e^{-i w \psi(x)}$ in powers of $\psi(x)$ and notice that when $\psi(x)$ is a power-law function of $x$, each integral in the series expansion can be performed analytically. In the case of a gSIS lens we have
\begin{align}\label{eq:gsis_wo_exact}
    F(w) 
    &= 
    -i w e^{\frac{i}{2} w y^{2} } 
    \sum_{n=0}^{\infty} 
    \frac{1}{n!}
    \left( \frac{-iw}{2-k} \right)^n  
    \nn \\
    &
    \times
    \int_0^{+\infty} {\rm d}x \,x^{1+n(2-k)} J_0(wxy) e^{i w x^2/2} 
    \nn \\
    &=  
    \sum_{n=0}^{\infty} 
    \frac{\Gamma \left(\frac{n}{2 A}+1\right) }{n!}
    \left[2^{\frac{1}{2A}} A (-i w)^{k/2} \right]^n
    L_{\frac{n}{2A}}\left( i w y^2/2\right)\;,
\end{align}
where once again $A = 1/(2-k)$. Notice that for low $w$, this expansion reduces to our approximation obtained in Eq.~\eqref{eq:low-w_gSIS}.
Moreover, in the particular case of $k = 1$ (SIS) we re-obtain the series representation first derived in \cite{Matsunaga:2006uc}:
\begin{equation}\label{eq:sis_wo_exact}
\begin{aligned}
    F(w)
    &=
    e^{\frac{i}{2} w y^{2}} 
    \sum_{n=0}^{\infty} \frac{\Gamma\left(1+\frac{n}{2}\right)}{n !} \\
    &\times\left(2 w e^{i \frac{3 \pi}{2}}\right)^{n / 2}{ }_{1} F_{1}\left(1+\frac{n}{2}, 1 ,-\frac{i}{2} w y^{2}\right)\;.
\end{aligned}
\end{equation}

In order to recover the oscillatory features of $F(w)$ accurately for large-enough $w$, the series in \eqref{eq:gsis_wo_exact} and \eqref{eq:sis_wo_exact} need to be truncated at relatively high values of $n$.
In our comparisons this truncation is made after reaching a $10^{-15}$ precision. As an example, for $w = 200$ and $y = 0.3$ this is obtained at around $n = 740$. For lower $w$ at fixed $y$, convergence is reached at lower $n$.
As a result, the features due to the GO, bGO and rGO terms are not manifest from this series expansion.

The series \eqref{eq:gsis_wo_exact} is impractical for most applications, since the evaluation of the Laguerre polynomials is slow. Moreover, for imaginary argument they grow exponentially in $n$. Therefore, many terms in the series are required to reach convergence, even at moderate $w$. Thus, as we will see, these results are outcompeted by numerical implementations of the diffraction integral in practical applications.
Additionally, this series expansion is difficult to generalize to other lens models such as the CIS lens, since in that case we do not have a closed-form solution for the integrals.

\section{Regularized Contour Flow}\label{sec:wo_contour}

\begin{figure*}[t]
\includegraphics[width=0.5\textwidth]{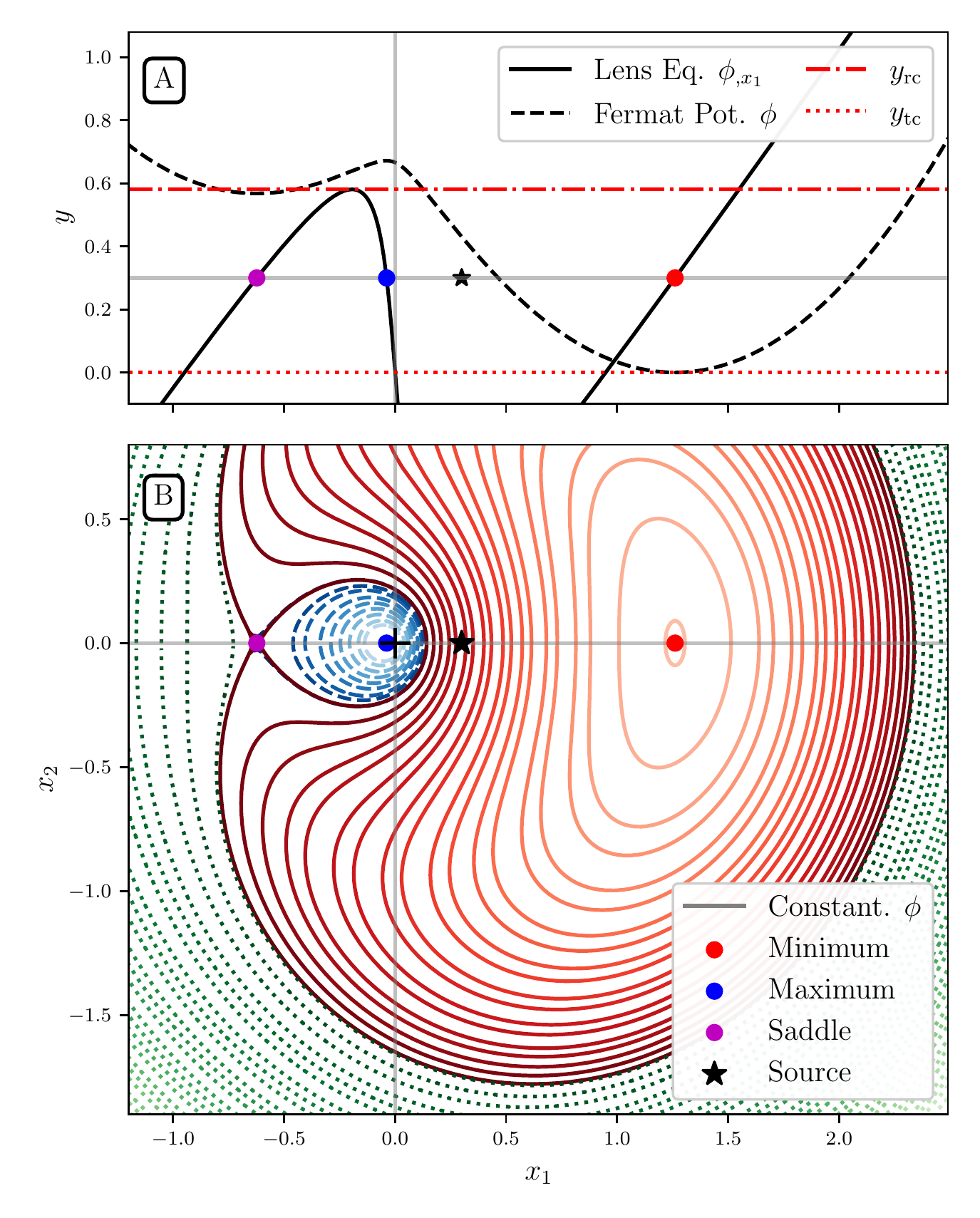}%
\includegraphics[width=0.5\textwidth]{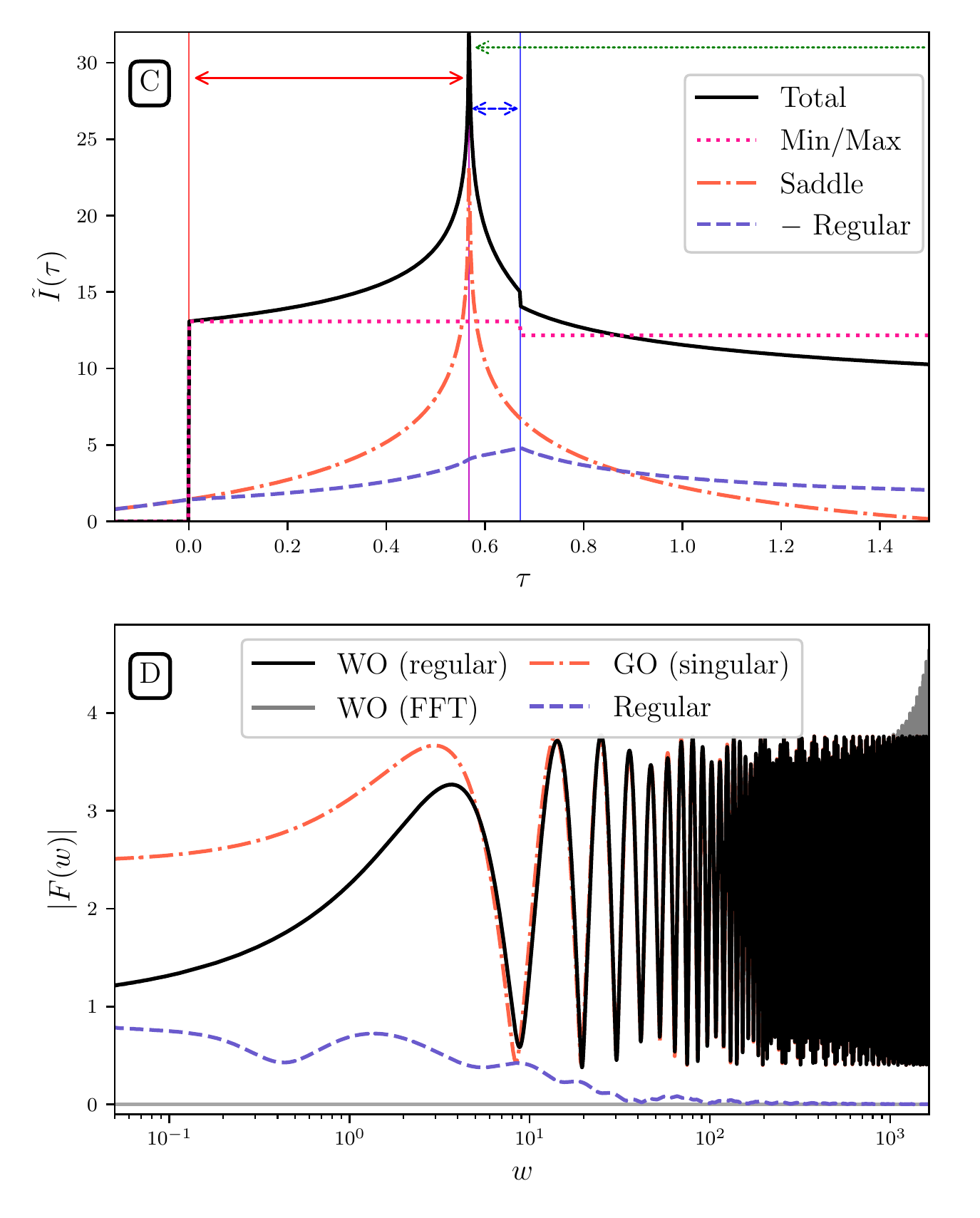}
\caption{WO lensing via contour flow with regularization. The lens is a cored isothermal sphere (see Sec.~\ref{sec:lensing_wo} and Tab.~\ref{tab:lenses_summary}) with a core size $x_c=0.05$, impact parameter $y=0.3$ and zero external convergence and shear.
\textbf{Panel A:} Images are found across the lens-centre-source axis by solving the lens equation. Images are located at stationary points of the Fermat potential $\phi$ (dashed line). Caustics are shown in red.
\textbf{Panel B}: The time-domain integral $\tilde I(\tau)$ (Eq.~\eqref{eq:contour_integrand}) is computed across equal-time contours in the lens plane. The contours start at infinity (dotted), maximum (dashed) and minimum of $\phi$ (solid) and are evolved until they approach the saddle point (purple dot).
\textbf{Panel C}: $\tilde I(\tau)$ is computed by summing the contributions from all contours. Horizontal arrows show the extent of each contour from panel B, ending at $\tau\to\phi_J$ (vertical lines). The singular contributions from maxima/minima (dotted) and the saddle point (dash-dotted) are removed, leaving the regular part $\tilde I_{\rm reg}(\tau)$ (dashed).
\textbf{Panel D}: WO amplification factor. The total $F(w)$ (solid, Eq.~\eqref{eq:contour_F_split}) is the sum of a regular piece (dashed), obtained by Fourier-transforming $\tilde I_{\rm reg}(\tau)$, and GO contributions for all the images (dash-dotted). Fourier-transforming the total $\tilde I(\tau)$ without splitting regular and singular contributions introduces numerical error at high frequencies (solid gray).
\label{fig:lensing_singular_splitting}
}
\end{figure*}

We will now turn to general methods to numerically solve Eq.~\eqref{eq:lensing_wave optics}.
This Section presents the regularized contour flow, a calculation performed by Fourier transforming the integral in Eq.~\eqref{eq:lensing_wave optics} and evaluating the resulting time-domain integral on contours of equal time delay. Each set of contours is then flowed adaptively to a different value of the time delay until ``hitting'' a critical point, when the contour ends (several contours end at saddle points). The total integral is transformed back to frequency space by means of a fast-Fourier transform (FFT), after splitting the result into regular, smooth and singular contributions, associated to GO results. We follow Ulmer \& Goodman \cite{Ulmer:1994ij} (see also \cite{nakamura1999wave}) but use a different regularization used for the saddle points. The steps in the method are described in Fig.~\ref{fig:lensing_singular_splitting}, the precision parameters and their default values in Table \ref{tab:prec_param_contour}.

\subsection{Adaptive Sampling in the Time Domain}

We will first compute the amplification factor in time-delay space. 
Fourier-transforming the integrand in Eq.~\eqref{eq:lensing_wave optics} yields
\begin{eqnarray}
  \tilde I(\tau) 
  &=& 
  \frac{1}{2\pi} 
  \int \de^2x \int_{-\infty}^{+\infty} \de w\,
  e^{i w\left(\phi(\vect x, \vect y)-\tau\right)}  
  \nonumber \\
  &=&  
  \int \de^2x \, 
  \delta_D \left( \phi(\vect x,\vect y) - \tau\right) \,,
 \label{eq:lensing_contour_time_integral}
\end{eqnarray}
where $\delta_D(x)$ is the Dirac-delta function.
This expression is the primitive of the Green's function: the time-domain lensed waveform is given by $h(t) = \int \de \tau \, \de \tilde I  (\tau-t) / \de \tau  \, h_0(\tau)$.
One can compute the function $\tilde I(\tau)$ by binning in time delays, cf.~Ref.~\cite{Diego:2019lcd, Cheung:2020okf}.
Instead, in this approach we are going to reduce the 2D integral into a 1D integral over the contours where the argument of the Dirac delta is zero:
\begin{equation}\label{eq:contour_integrand}
    \tilde I(\tau) 
    =
    \sum_k \oint_{\gamma_k}
    \frac{\de s}{|\vect \nabla_{\vect x}\phi(\vect x(\tau,s),\vect y)|}
    \,,
\end{equation}
where $s$ is the arc-length distance that parametrizes the $k$-th contour $\gamma_k$ of constant $\tau$ (it is possible to reparameterize the contour so the integrand in Eq.~\eqref{eq:contour_integrand} reduces to $\int\de u$, cf.~Ref.~\cite[Eq.~9]{Ulmer:1994ij}). The sum is over all contours contributing to a given $\tau$ (see below). 

Contours of constant Fermat potential are orthogonal to the gradient of $\phi$, hence given by
\begin{equation}
    \frac{\partial \vect x(\tau, s)}{\partial  s} \cdot \vect \nabla_{\vect x}\phi(\vect x, \vect y) = 0\,.
\end{equation}
To sample $\tilde I(\tau)$, we flow the contours as
\begin{equation}\label{eq:contour_flow}
    \vect x(\tau+\Delta \tau, s) 
    \simeq 
    \vect x(\tau, s) 
    + \Delta\tau \frac{\vect\nabla_{\vect x}\phi}{|\vect\nabla_{\vect x}\phi|^2}\,,
\end{equation}
i.e.~each point is displaced (linearly) in the direction of the Fermat potential's gradient, with $\Delta\tau$ positive/negative if $\tau$ is increasing/decreasing. 
The integral $\tilde I(\tau)$ is sampled adaptively on each family of contours. We choose the step $\Delta \tau$ depending on the rate of variation $\tilde I$, according to the following prescription
\begin{equation}\label{eq:contour_flow_steps}
    \Delta\tau_{i+1} = \min\Big(\eta \frac{|\de\tilde I/\de\tau|}{|\de^2\tilde I/\de\tau^2|}, \eta^\prime\Delta\tau_i, \Big)\,.
\end{equation}
Here $\eta,\eta^\prime$ are precision parameters and $\de \tilde I/\de\tau, \de^2\tilde I/\de\tau^2$ are computed numerically from the previous iterations. The variation $\Delta\tau$ is kept within a minimum and maximum values. The algorithm stops if a prescribed number of iterations is reached.

At each step $\tau_i$ we refine the contour so the distance between its nodes is small compared to the local curvature radius of the contour (see Table \ref{tab:prec_param_contour}).
Contours are flown until reaching a critical point, where they either shrink to a point (maxima, minima and cusps) or become non-differentiable (saddle points). 
Therefore, we first find the critical points $\vect x_J$ such that $\vect \nabla_{\vect x}\phi(\vect x_J,\vect y)=0$ (i.e.~GO images), Eq.~\eqref{eq:lens_eq}. Since we work with symmetric lenses we search only along the lens-source direction $x_1\equiv \vect x\cdot\vect y / y$ (Fig.~\ref{fig:lensing_singular_splitting} panels A, B).

\begin{table}[]
    \centering
    \begin{tabular}{c c  p{0.6\columnwidth}}
        Parameter         & \;Value\; & Description \\ \hline \\[-10pt]
        \multicolumn{3}{l}{Contour initialization:} \\
        \texttt{r\_out}     & 30 & outer contour radius \\
        \texttt{r\_in}      & $10^{-3}$ & contour radius at min/max \\
        \texttt{nodes\_ini}     & 500 & initial nodes in contour \\ \hline \\[-10pt]
        \multicolumn{3}{l}{Contour adaptive flow, Eq.~\eqref{eq:contour_flow}:} \\
        \texttt{dtau\_0}     & $10^{-4}$ & initial $\Delta \tau$ \\ 
        $\eta$     & $0.1$ & \multirow{2}{*}{adaptive stepsize, Eq.~\eqref{eq:contour_flow_steps}
        }  \\ 
        $\eta^\prime$     & $1.5$ &  \\ 
        \texttt{dtau\_min}     & $10^{-4}$ & minimum variation \\ 
        \texttt{dtau\_max}     & $0.1$ & maximum variation \\ 
        \texttt{max\_steps}      & $10^{4}$ & maximum number of steps \\ 
        \hline \\[-10pt]
        \multicolumn{3}{l}{Contour node refinement:} \\
        $\delta_{\rm max}$     & $0.02$ & insert new node if $d_{k-1,k}>\delta_{\rm max}R_k$ \\ 
        $\delta_{\rm min}$     & $0.005$ & remove $k$ node if $d_{k-1,k+1}<\delta_{\rm min}R_k$ \\ 
        \texttt{nodes\_min}    & $10$ & minimum nodes in contour \\ 
       \hline \\[-10pt]
        \multicolumn{3}{l}{FFT parameters:} \\
        $\tau_{\rm max}$  & 500   & maximum sampled $\tau$ \\
        $N_{\rm FFT}$     & $2^{17}$ & FFT points \\
        \texttt{tau\_min\_extend} & $-0.1$ & added to minimum sampled $\tau$ \\
        \hline \\[-10pt]
        \multicolumn{3}{l}{Regularization:} \\
        \texttt{window\_width}    & 0.2 & window function width (relative)  \\ 
        $T$               & 30 & saddle point width, Eq.~\eqref{eq:saddle_point_window} \\\hline
    \end{tabular}
    \caption{Summary of the precision parameters used for the contour method. Here $d_{l,m}$ is the distance between the $l$,$m$-th node points and $R_l$ is the local curvature radius of the contour evaluated at the $l$-th node.}
    \label{tab:prec_param_contour}
\end{table}

For a given value of $\tau$ there can be zero, one or multiple contours depending on the lens configuration. 
The total number of contours changes when $\tau$ crosses the values of the time delay $\phi_J$ associated to critical points and the integrand in Eq.~\eqref{eq:contour_flow} becomes singular. Hence, critical points are associated with singularities and discontinuities of $\tilde I(\tau)$.
Contours do also end in non-regular points of the lensing potential, which are not associated to GO images.%
\footnote{If $\phi$ is made smooth a GO image forms, but it satisfies $\mu \to 0$ in the singular limit (e.g.~for the CIS \cite{our_paper}).}

In the weak-lensing regime there is a single type I image at the minimum of $\phi$: the contours then flow between the minimum and infinity, where they approach circles/ellipses centered around the source.
Strong lenses produce multiple contours in certain ranges of $\tau$. For the SIS, gSIS and CIS in the strong-lensing regime there are three regions to consider: 1) from infinity down to the saddle point, 2) from the minimum up to the saddle point and 3) from the maximum down to the saddle point.
Contours that begin near minimum/maximum are initiated with a small radius around the critical points. The contour that asymptotes to infinity is initiated at a large radius around the source (see Table \ref{tab:prec_param_contour}).
These regions and their contribution to $\tilde I(\tau)$ are shown in Fig.~\ref{fig:lensing_singular_splitting} (panels B and C).

\subsection{Time-domain Regularization and GO Counterterms}
\label{subsec:time_dom_reg}

The frequency-domain amplification factor can be computed from Eq.~\eqref{eq:lensing_contour_time_integral} via inverse Fourier transform. We will perform this operation via fast-Fourier transform, for which we interpolate $\tilde I(\tau)$ on an equally spaced grid with 
$N_{\rm FFT}$
points and spacing $\delta\tau$, whose range and density are determined by the minimum and maximum values of $w$ that we are interested in. 
In order to avoid boundary effects, we apply a Tukey window function to the time-domain signal. We extend $\tau$ towards negative values, such that the window function is $\simeq 1$ over all $\phi_I$ corresponding to GO images.

One difficulty is dealing with the discontinuities and singularities in $\tilde I(\tau)$ associated to GO images, as they produce components at arbitrarily high frequency. The discretization needed for the FFT causes aliasing of the frequencies $w>w_{\rm max}$, contaminating the computed signal at high $w$ (Fig.~\ref{fig:lensing_singular_splitting}, panel D). 

To avoid numerical artefacts it is convenient to treat the contribution of stationary points separately \cite{Ulmer:1994ij}. Hence we split the integral into a regular and a singular part
\begin{equation}\label{eq:contour_split}
    \tilde I(\tau)
    =
     \tilde I_{\rm reg}(\tau) 
     +
     \sum_J\tilde I^J_{\rm sing}(\tau)
    \,.
\end{equation}
The l.h.s.~is obtained by evaluating Eq.~\eqref{eq:contour_integrand} numerically, as described above. The singular contributions have closed-form expressions, which we give below. 
The WO amplification factor follows by Fourier transforming
\begin{equation}\label{eq:contour_F_split}
    F(w) 
    = 
    F_{\rm reg}(w) + \sum_J F_{\rm sing}^{J}(w)
    \,.
\end{equation}
The regular contribution is the FFT from $\tilde I_{\rm reg}(\tau)$. The singular contributions are the GO amplification factors for each image, Eq.~\eqref{eq:lensing_geometric_optics}, or related to them. 
Note that the above splitting is arbitrary and valid as long as the time and frequency domain terms are consistent. Therefore, we can add any such terms in order to make the computation more robust.
We will now discuss these terms for different types of critical points.

Type I/III images (minima/maxima of the Fermat potential) correspond to discontinuities in $\tilde I(\tau)$. In this case
\begin{equation}
    \tilde I_{\rm sing}^M(\tau) 
    = 
    2\pi |\mu_J|^{1/2}\, \theta(\pm (\tau - \phi_J))
    \,,
\end{equation}
where $\theta$ is the Heaviside step function and $+/-$ corresponds to a minimum/maximum with time delay $\phi_J$.
The corresponding GO contribution reads
\begin{equation}\label{eq:contour_extrema_freqdomain}
  F_{\rm sing}^M(w) 
  = 
  \mp |\mu_J|^{1/2}e^{i w \phi_J}
  \,.
\end{equation}
The discontinuity is interpreted as a family of contours ceasing to exist at the extremum. In the case of a cusp ($x\to0$ in the SIS and gSIS) the contour ceases to exist but no discontinuity forms because $\mu_J=0$.

A type II image (saddle point) with time delay $\phi_J$ produces a logarithmic divergence in $\tilde I(\tau)$ \cite{Ulmer:1994ij}
\begin{equation}\label{eq:saddle_contrib_timedomain_infty}
    \tilde I_{\rm sing}^S(\tau) 
    \simeq  
    -2 |\mu_J|^{1/2}\log|\tau -\phi_J| + C
    \,,
\end{equation}
with $C$ an integration constant.
This calculation assumes that the quadratic approximation around the saddle-point, $\phi \simeq (\Delta x_1/a)^2- (\Delta x_2/b)^2$, is valid for arbitrarily large separations $\Delta\vect x$.%
\footnote{
Including only a finite region around the saddle point yields
\begin{equation}\label{eq:saddle_contrib_timedomain_exact}
    \tilde I_{\rm sing}^S(\tau) 
    =
    4 |\mu_J|^{1/2}\cosh^{-1}\left(\sqrt{\frac{\delta\tau}{|\tau -\phi_J|}}\right)\,,
\end{equation}
where $\delta\tau$ defines the limit of the contour around the saddle point. 
Equation \eqref{eq:saddle_contrib_timedomain_infty} follows when $\frac{\delta\tau}{|\tau -\phi_J|}\gg 1$.
We do not use the more accurate Eq.~\eqref{eq:saddle_contrib_timedomain_exact}, as it does not have a closed-form Fourier transform.
}
The corresponding GO contribution reads
\begin{equation}\label{eq:saddle_contrib_infty}
  F_{\rm sing}^S(w) = i |\mu_J|^{1/2}e^{iw \phi_J}\,,
\end{equation}
This term contributes at arbitrarily large values of $\tau$, causing spurious low-frequency behaviour upon FFT. 
One can avoid this issue by windowing the singular contribution $\tilde I_{\rm sing}^S(\tau)\to W(\tau,T)\tilde I_{\rm sing}^S(\tau)$ in Eq.~\eqref{eq:saddle_contrib_timedomain_infty}. 
Choosing 
\begin{equation}\label{eq:saddle_point_window}
W(\tau, T) = e^{-|\tau-\phi_J|/T}    
\end{equation}
preserves the singular behaviour and avoids the low frequency problems if $\delta \tau \ll T \ll \delta\tau N_{\rm FFT}$, where $N_{\rm FFT}$ is number of sampled points.

This choice produces a closed-form expression for the frequency domain
\begin{equation}
    F_{\rm sing}^{S}(w,T) 
    =
    \frac{- i w}{\pi}|\mu_J|^{1/2}e^{iw \phi_J}
    \left(\mathcal{I}_+ + \mathcal{I}_-\right)
    \,,
\end{equation}
where 
\begin{eqnarray}
    \mathcal{I}_{\pm}(w,T) &\equiv& \int_0^\infty \de t \, \log(t)\, e^{- t/T \pm iw t} 
    \nonumber \\ 
    &=& 
    \frac{-i T}{i\pm wT}\left(\gamma_E + \log\left(T^{-1 }\mp i w\right)\right)
    \,.
\end{eqnarray}
Note that $\lim_{T \to \infty} \mathcal{I}_{\pm} = \mp\frac{i}{w}\left(\gamma_E + \log(w) \mp i\frac{\pi}{2}\right)$, recovering Eq.~\eqref{eq:saddle_contrib_infty}. 

Panel C of Fig.~\ref{fig:lensing_singular_splitting} shows the integral in the time domain, including both the regular and the different singular contributions.
Panel D shows the different contributions to $F(w)$: at low frequencies the singular part (corresponding to GO predictions) is compensated by the regular contribution and recovers the low-frequency limit $F(w\to0)\to 1$.
Without splitting the singular part, $F(w)$ loses precision at high frequencies and eventually becomes unreliable (solid gray line).

After the regularization, discontinuities remain in derivatives of $\tilde I(\tau)$ at the critical points $\tau=\phi_J$. When Fourier transformed, discontinuous $\frac{\de}{\de\tau}\tilde I(\tau)$ corresponds to corrections $\propto 1/w$ in the amplification factor. These terms are precisely the bGO and cusp contributions discussed in Sec.~\ref{sec:lensing_go}. Discontinuities extend to any derivative $\frac{\de^n}{\de\tau^n}\tilde I(\tau)$, with corresponding corrections $\Delta F^{(n)}\propto w^{-n}$ as higher order bGO terms.
Numerically, the discontinuous derivatives cause aliasing in the FFT and are a source of error, although the scaling with $w$ makes these terms subdominant in the computation of $F(w)$ (see discussion in Sec.~\ref{sec:compare_point}).
Eventually, our regularization method could be extended to split discontinuities in derivatives of $\tilde I(\tau)$ as higher precision is required.

Finally, let us mention that computing derivatives of the amplification factor accurately requires handling discontinuities in $\frac{\de}{\de \tau}\tilde I(\tau)$. This is because an additional derivative (e.g.~with respect to the lens parameter $\Theta$) promotes the $w^{n}$ terms in $F$ to $\sim w^{n+1}$ in $\de F/\de \Theta$: our regularization removes $\sim w$ terms, but leaves terms $\sim w^0$, which contribute significantly to aliasing.
As computing derivatives of $F(w)$ is important in some applications of WO lensing (e.g.~Fisher-matrix forecasts \cite{our_paper}), we discuss a workaround in Appendix \ref{sec:appendix_F_derivs}.

\section{Complex deformation}\label{sec:wo_complex_def}

\begin{figure*}[t]
 \includegraphics[width=0.49\textwidth]{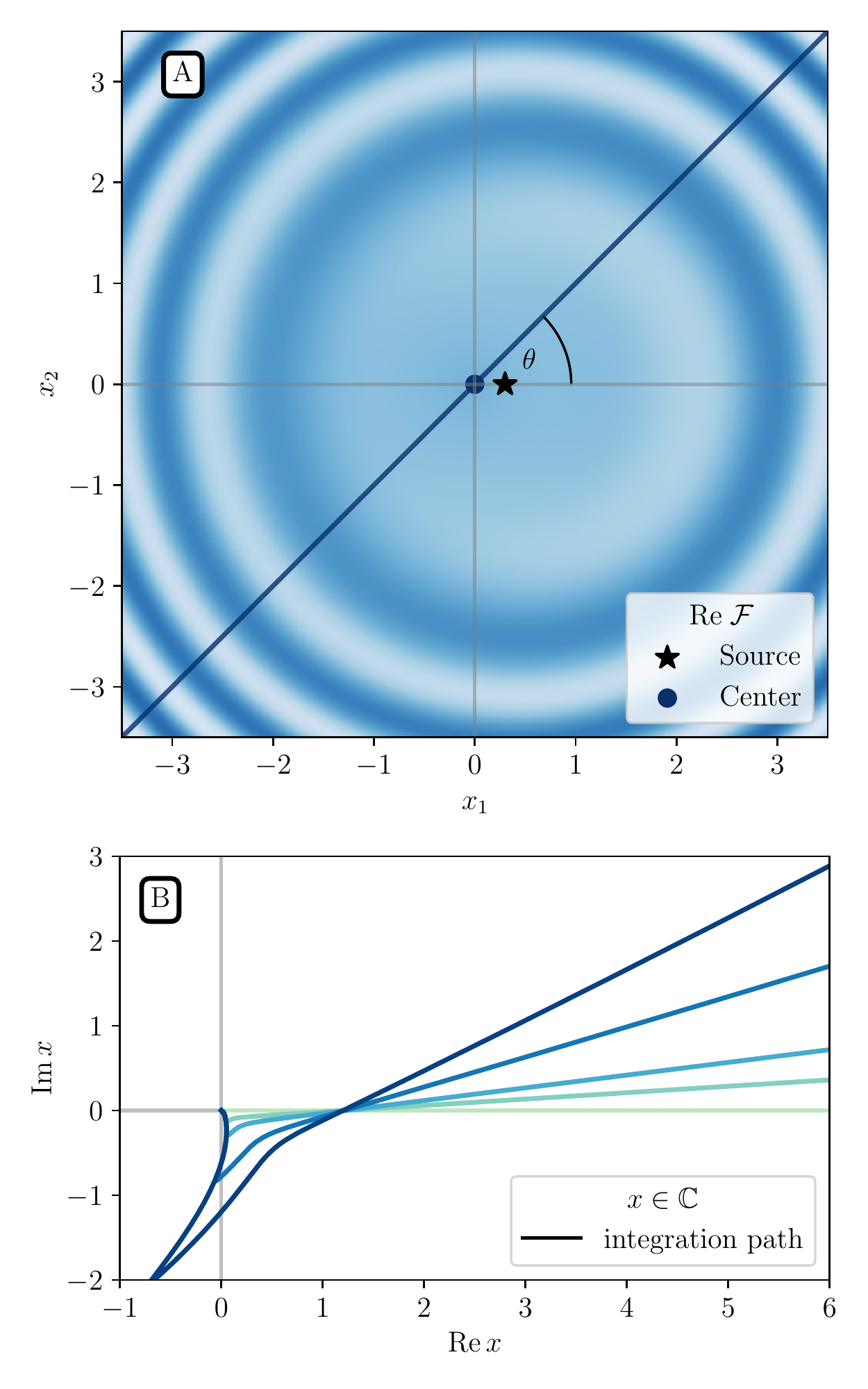}
 \includegraphics[width=0.49\textwidth]{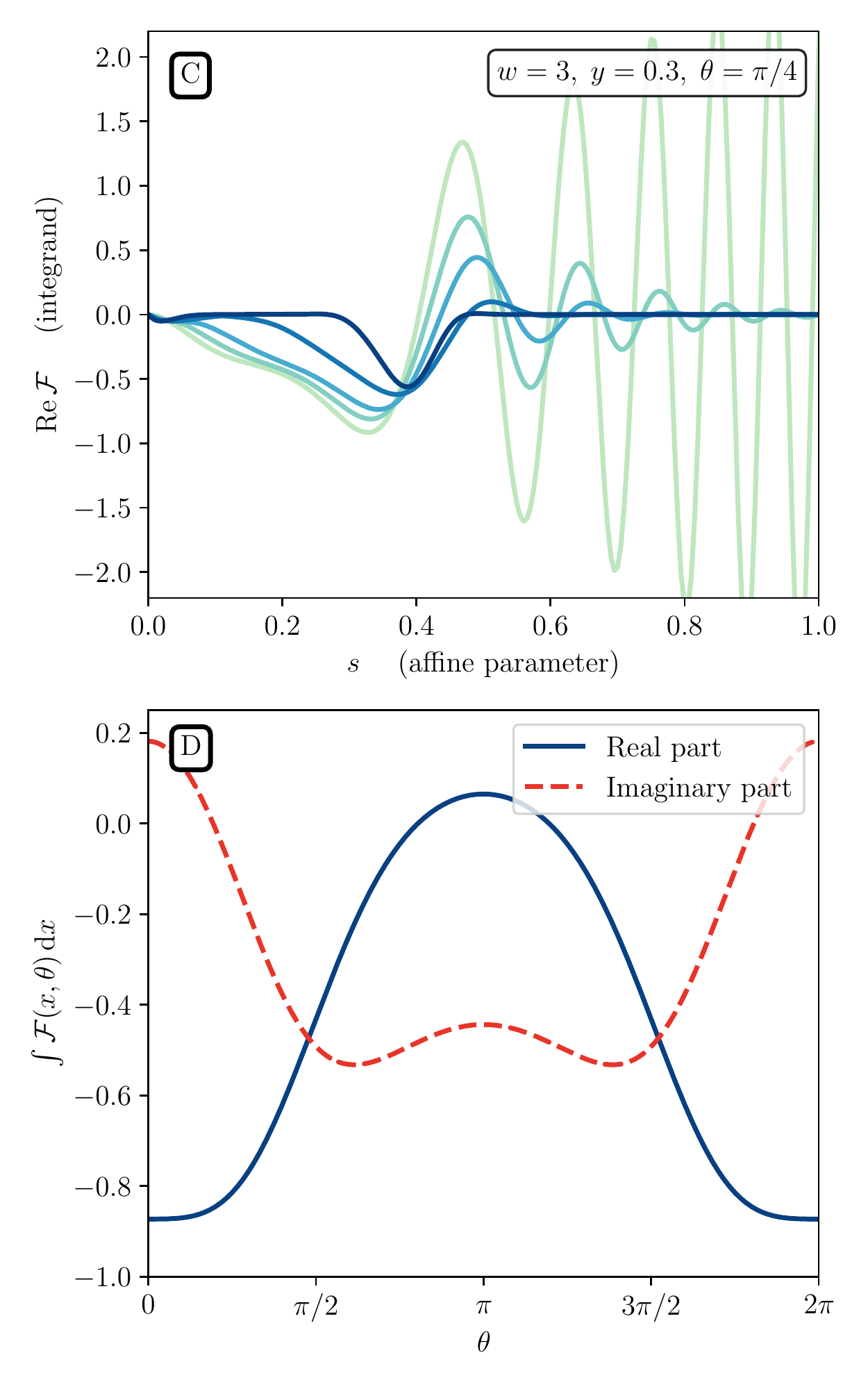}
 \caption{\label{fig:complex_flow}WO lensing via complex-deformation method. The lens is a SIS with $y=0.3$ and fixed $w = 3$. 
 \textbf{Panel A}: The integrand $\mathcal F$ in Eq.~\eqref{eq:complex_method_angular_int} (showed in the density plot) is highly oscillatory in the lens plane.
 First, we select a value for the angle $\theta$ in the lens plane (blue line) where the flow equation is then applied, for this example we choose $\theta = \pi/4$. 
 \textbf{Panel B}: Evolution of the integration path under the flow equation \eqref{eq:flow_eq_rescaled}, starting from the positive real line $x>0$ (light green line). Darker lines are obtained by increasing $\lambda$ in the flow.
 \textbf{Panel C}: The integrand at fixed $\theta$ evaluated on the paths of Panel B (as an example, only the real part of $\mathcal F$ is shown). We use the affine parameter $s$, which parametrizes each path, as the $x$ axis. The initial integration contour $\lambda = 0$ gives large oscillations that are difficult to treat numerically. Very quickly for $\lambda > 0$ the integrand becomes exponentially damped at large $s$. 
 \textbf{Panel D}: Integral over the complex variable $x$ as a function of the angle $\theta$, obtained by repeating the steps from Panel A to C for different $\theta$s. The amplification factor in Eq.~\eqref{eq:complex_method_angular_int} is then obtained by integrating over $\theta$: the latter integration has no convergence problems.
 }
\end{figure*}

Another method to improve the convergence of the diffraction integral \eqref{eq:lensing_wave optics} relies on analytically continuing the integration variable into the complex plane.
We follow a variation of the procedure described in \cite{Feldbrugge:2019fjs} (see also \cite{Feldbrugge:2020ycp,Feldbrugge:2020tti,Suvorov:2021uvd}).
Initially, we briefly review how to obtain a good integration contour in the complex plane for highly-oscillatory 1D integrals.
Then, we generalize the discussion to the more relevant case of 2D diffraction integrals with non-analytic features, as appear in gravitational lensing.

\subsection{Flow of the integration domain}

To outline the method, we start with a prototypical oscillatory integral, which resembles \eqref{eq:lensing_wave optics} in one dimension
\begin{equation}\label{eq:diff_integral_1d}
    F(w) 
    = 
    \int_{-\infty}^{+\infty} \de x 
    \, 
    g(x) e^{i w f(x)}
    \;,
\end{equation}
where $f(x)$ and $g(x)$ are assumed to be analytic in the full complex plane $z = x + i y \in \mathbb C$ (later we will mention how to deal with possible branch-cuts or poles away from the real line).\footnote{The distinction between $f(x)$ and $g(x)$ is ambiguous in Eq.~\eqref{eq:diff_integral_1d}. To partially fix the ambiguity, we assume that the convergence properties of the integral at infinity are solely determined by $f(x)$. As we are going to elaborate, convergence is established by deforming the integration contour into the complex plane (in particular, by tilting the contour above or below the real line, depending on the properties of $f(x)$).} Clearly, the limit of large $w$ corresponds to the saddle-point, or GO, approximation for Eq.~\eqref{eq:diff_integral_1d} (when dealing with complex functions, critical points can only be saddles).

First, let us recall how to make sense of this type of oscillatory integrals, which are not manifestly convergent (see e.g.~\cite{Serone:2017nmd} for more details on this procedure).
The integral \eqref{eq:diff_integral_1d} can be extended to the complex plane $z = x + i y$ by deforming the contour of integration away from the real line (because of Cauchy's theorem the final answer does not change). In particular, the contour can be decomposed into so-called steepest-descent paths, each associated with a saddle point. (Some saddles might be irrelevant in this decomposition and neglected, as they are not encountered when moving the real-line integration contour into the complex plane).  
The steepest-descent path $z(\lambda)$ of the function $f(x)$ associated with an isolated and non-degenerate saddle $z_c$ (i.e.~$f''(z_c) \neq  0$) satisfies the flow equation
\begin{equation}\label{eq:flow_eq_complex_deformation}
    \frac{\de}{\de \lambda}z(\lambda) 
    =
    i \frac{\partial f^* }{\partial z^*}
    \;,
\end{equation}
where $\lambda \in \mathbb{R}$ parametrizes the path and $z^*$ is the complex conjugate of $z$. From the equation above, it follows that
\begin{equation}
    i \frac\de{\rm d \lambda} f 
    =
    i \frac{\partial f}{\partial z} 
    \frac{{\rm d } z}{{\rm d \lambda}} 
    =
    - \left| \frac{\partial f}{\partial z} \right| ^{2}
    \;.
\end{equation}
Therefore, the real part of the exponent of the integrand \eqref{eq:diff_integral_1d} monotonically decreases along the steepest-descent path: the integral is now manifestly convergent, as the integrand decays exponentially.\footnote{Together with the steepest-descent path, each saddle has an associated steepest-ascent path. We do not discuss these paths here, as they are unimportant for our discussion.} On the other hand, the imaginary part of the exponent remains constant along the curve $z(\lambda)$. Additionally, the saddle point $z_c$ is reached only asymptotically in the flow (at $z_c$ the flow in Eq.~\eqref{eq:flow_eq_complex_deformation} stops, as $f'(z_c) = 0$).

The flow equation \eqref{eq:flow_eq_complex_deformation} can also be directly used to \emph{determine} the integration contour, without needing to identify the relevant saddle points beforehand. This procedure, introduced in \cite{Feldbrugge:2019fjs} and discussed below, is what we will use in our applications. 
We start by considering again Eq.~\eqref{eq:flow_eq_complex_deformation}. In this flow equation now we impose as initial condition at $\lambda = 0$ that $z(0) = x$, where $x$ is a generic point on the initial integration contour $\mathcal C_{\lambda = 0}$ (in the case of Eq.~\eqref{eq:diff_integral_1d}, $\mathcal C_{\lambda = 0} = \mathbb R$). Let us call $\mathcal C_{\lambda}$ the set of points $z(\lambda)$ at the ``time'' $\lambda$. 
As a result of Morse theory, the contour $\mathcal C_{\lambda}$ for $\lambda \rightarrow \infty$ converges to a steepest-descent contour (Eq.~\eqref{eq:flow_eq_complex_deformation} represents a smooth deformation of the initial path $\mathcal C_{\lambda = 0}$).
Therefore, just by repeat use of Eq.~\eqref{eq:flow_eq_complex_deformation} for all the points of the initial domain of integration, we can write the original integral for large enough $\lambda$ as
\begin{equation}\label{eq:complex_deformation_C_integral}
F(w) = \int_{\mathcal C_{\lambda}} \de z \, g(z) e^{i w f(z)}\;,
\end{equation}
where the exponent is now real (up to a $z$-independent imaginary part) and not oscillatory. 
Standard numerical techniques can now be applied to this integral.
Notice that $\mathcal C_{\lambda}$ does not depend on $w$
, therefore the path needs to be computed only once.
This property applies to GW lensing, where $w$ appears only linearly at the exponent. In situation where the Fermat potential depends on the frequency, as for wave diffraction in dispersive media, this is no longer the case.

\subsection{Extension to realistic lenses}

The procedure just outlined can be generalized to higher dimensions and applies to the diffraction integral \eqref{eq:lensing_wave optics}. For our practical purposes, however, it is simpler to reduce $F(w)$ to a set of one-dimensional integrals and apply the procedure above. To achieve this, we first write the integral over $\vect x$ in polar coordinates $x$ and $\theta$ as
\begin{equation}\label{eq:complex_method_angular_int}
    F(w) 
    =
    \frac{w}{2 \pi i} 
    \int_0^{2\pi} \de \theta \int_0^{+\infty} \de x \, 
    x \,e^{i w \phi(x, \theta, \vect y)}
    \;.
\end{equation}
The integral in $\theta$ is over a finite range and can be performed with standard numerical techniques. On the other hand, the integral over $x$ is highly oscillatory and is suitable for the analytic continuation procedure. The contour-deformation method has to be applied for various values of $\theta$, until the sampling of points is dense enough to guarantee numerical convergence.

The only difference with Eq.~\eqref{eq:diff_integral_1d} is the lower limit of integration, which stops as $x = 0$. In order to apply Cauchy's theorem, the initial and final contours $\mathcal C_{\lambda = 0}$ and $\mathcal C_{\lambda}$ must close (up to a semicircle at large $|x|$, which is negligible). However, if the point $z(0) = 0$ is evolved according to the flow, it will in general move away from the origin, thus leaving the sum of the initial and final contours open. To avoid this issue, we decide to alter the flow equation \eqref{eq:flow_eq_complex_deformation} in such a way as to force the points close to the origin not to evolve. To do so we rescale the flow variable $\lambda \rightarrow h_{z_0} \lambda$, where $h_{z_0}$ is a function that depends on the initial position $z_0 \equiv z(0)$. Note that changing the flow equation does not modify the final result due to Cauchy's theorem. The flow equation then becomes
\begin{equation}\label{eq:flow_eq_rescaled}
    \frac{\de}{\de \lambda}z (\lambda) 
    =
    i \frac{\partial f^* }{\partial z^*} h_{z_0}
    \;.
\end{equation}
For convenience we choose $h_{z_0}$ to be $h_{z_0} = \theta_\varepsilon(z_0 - \delta)$, where $\theta_\varepsilon(x) \equiv \frac{1}{2}[\tanh( x /\varepsilon ) + 1]$ is smooth, and converges to the Heaviside step function for $\varepsilon \to 0$. The function $h_{z_0}$ interpolates between $h_{z_0} \simeq 0$ for points $z_0 < \delta$ and $h_{z_0} \simeq 1$ for $z_0 > \delta$.
The parameter $\varepsilon$ sets how sharp the transition is. 
In the application of the following sections we set $\varepsilon = 10^{-3}$ and $\delta = 10^{-2}$.

An additional complication arises when considering realistic lenses, which typically feature non-analytic lensing potentials (even the simplest example, the point lens $\psi = \log x$, has a branch-cut 
). Fortunately, our procedure is not significantly altered as long as we are dealing with branch cuts (and possibly poles, but we are not going to encounter them in our lensing models). 
Away from the branch cut the exponent is analytic, and the flow equation can be applied without modifications. It is possible however that some points are driven towards the branch cut during the flow. In such cases, to avoid them crossing the cut, we decide to stop the flow. This can be implemented case by case (depending on the location of the cuts) by modifying the flow. 
Thus, we multiply the right-hand side of Eq.~\eqref{eq:flow_eq_rescaled} by a function $b(z(\lambda))$ with the requirement that for $z(\lambda)$ approaching the cut, $b(z(\lambda)) \rightarrow 0$ smoothly, while $b(z(\lambda))\simeq 1$ everywhere else.
For the case of a point lens we have a branch cut for ${\rm Re}\, z < 0$. 
Following the same logic as for $h_{z_{0}}$, with this lens we choose $b(z) = 1 - \theta_\varepsilon(\varphi - \pi_\delta) - \theta_\varepsilon(-\varphi - \pi_\delta)$, where $z$ is written in polar coordinates $z = r e^{i \varphi}$ and we defined $\pi_\delta \equiv \pi - \delta$. With this choice, the evolution of the contours is halted as $\varphi$ approaches $\pm \pi$. For the function $b(z)$, in the next section we will use the values $\varepsilon = 1 / 200$ and $\delta = 10^{-1}$.

Since we are modifying the flow equation, some of the nice properties of Eq.~\eqref{eq:flow_eq_complex_deformation} are partially lost. In particular, it will no longer be true that the imaginary part remains constant along the final contour $\mathcal C_{\lambda}$ (hence some mild oscillations can reappear). In practice, for the cases we will consider, this is not an issue since the modifications only affect the contour close to the origin, while leaving the behaviour at large $|z|$ unaffected (the convergence properties of the integral are thus preserved).

We also notice that the integration contour at infinity converges to the 45-degrees line ${\rm arg}\, x  = \pi / 4$, ${\rm Re}\, x>0$ (for large $x$ the Fermat potential is dominated by its quadratic part). The contour is however deformed as $x$ approaches the origin, due the Lensing potential.
Therefore, the method outlined in this Section generalizes Eq.~\eqref{eq:gaussian_contour} and optimizes the choice for the contour.
 
Let us see how this procedure is applied in a particular lens model, the SIS lens. First, we fix a value for the angular variable $\theta$ in Eq.~\eqref{eq:complex_method_angular_int}. We take it to be $\theta=\pi/4$ in this example (Fig.~\ref{fig:complex_flow}, panel A). Then, we evolve the integration contour from the positive real line $x>0$ to the complex plane. Since this lens model does not introduce branch cuts or poles, the evolution of the path is obtained using Eq.~\eqref{eq:flow_eq_rescaled}. Then, it is stopped at some given flow time $\lambda = T$ and the path is truncated at some large value of $|x|$. The evolution of the paths is shown in Fig.~\ref{fig:complex_flow}, Panel B. We can notice that all the paths cross the real line at a fixed location: this corresponds to a saddle point of the exponent in Eq.~\eqref{eq:complex_method_angular_int} (this is however different from the saddles corresponding to the GO approximation, since we are fixing $\theta$ here).
Once the paths are obtained, we can evaluate the integrand in Eq.~\eqref{eq:complex_method_angular_int} (that we call $\mathcal F$) on each path. This quantity is now also a function of $w$. As an example, we show $\mathcal F$ for fixed $w$ and $\theta$ in Fig.~\ref{fig:complex_flow}, Panel C. Clearly, the problematic oscillations in $\mathcal F$ get damped very quickly as the flow progresses. For large values of $\lambda$ the integral becomes very localized around the saddle point.
Finally, one has to repeat the steps A-C for different values of $\theta$. The integral of $\mathcal F$ over $x$ as a function of $\theta$ is shown in Fig.~\ref{fig:complex_flow}, Panel D: for moderate $w$, this function is not oscillating too rapidly and can be integrated easily. 

In the results presented in the following sections, we use the following settings for the complex-deformation method.
The contours are evaluated for $n_1 = 25$ values of $\theta$, uniformly distributed between $0$ and $2 \pi$ (for symmetric lenses one can limit to the range $0$ to $\pi$). The integrand is then sampled over a larger number of values of $\theta$, $n_1 \cdot n_2$, with $n_2 = 25$. We can sidestep evaluating $n_1 \cdot n_2$ contour flows because the final integral is independent on the choice of the contour. Therefore, for a given angle, we use the contour evaluated with the nearest value of $\theta$.

For the flow of each contour, we sample the initial contour along $x > 0$ with $n_x = 340$ points, not uniformly distributed but concentrated towards $x = 0$.
The flow equation is implemented in \texttt{python}, using the \texttt{odeint} function of the \texttt{scipy} package on default settings.
After applying the flow equation, the final contour is interpolated over $n_x^{\rm interp} = 1000$ points. 

\section{Accuracy and performance} \label{sec:wo_compare}

We will now discuss the accuracy of the algorithms described in Sec.~\ref{sec:wo_contour} and \ref{sec:wo_complex_def} and their convergence to systematic expansions, Sec.~\ref{sec:lensing_expansions}. We first compare the results for a point lens (Sec.~\ref{sec:compare_point}) and then against each other (Sec.~\ref{sec:compare_methods}) for the other axially-symmetric lenses in Table \ref{tab:lenses_summary}. We end by discussing the performance of the different methods (Sec.~\ref{sec:compare_performance}). 
All the comparisons are made with impact parameter $y = 0.3$; similar conclusions are reached for different values of $y$, sufficiently far from caustics.
The precision parameter used in the contour method are summarized in Table \ref{tab:prec_param_contour}.

\subsection{Comparison to point lens}\label{sec:compare_point}

In order to assess the goodness of our numerical methods, we can compare to the point-lens model ($\psi(x) = \log x$), where the diffraction integral is known analytically (see Eq.~\eqref{eq:pt_lens_analytic}). 

\begin{figure}[t]
\begin{center}
  \includegraphics[width=0.98\columnwidth]{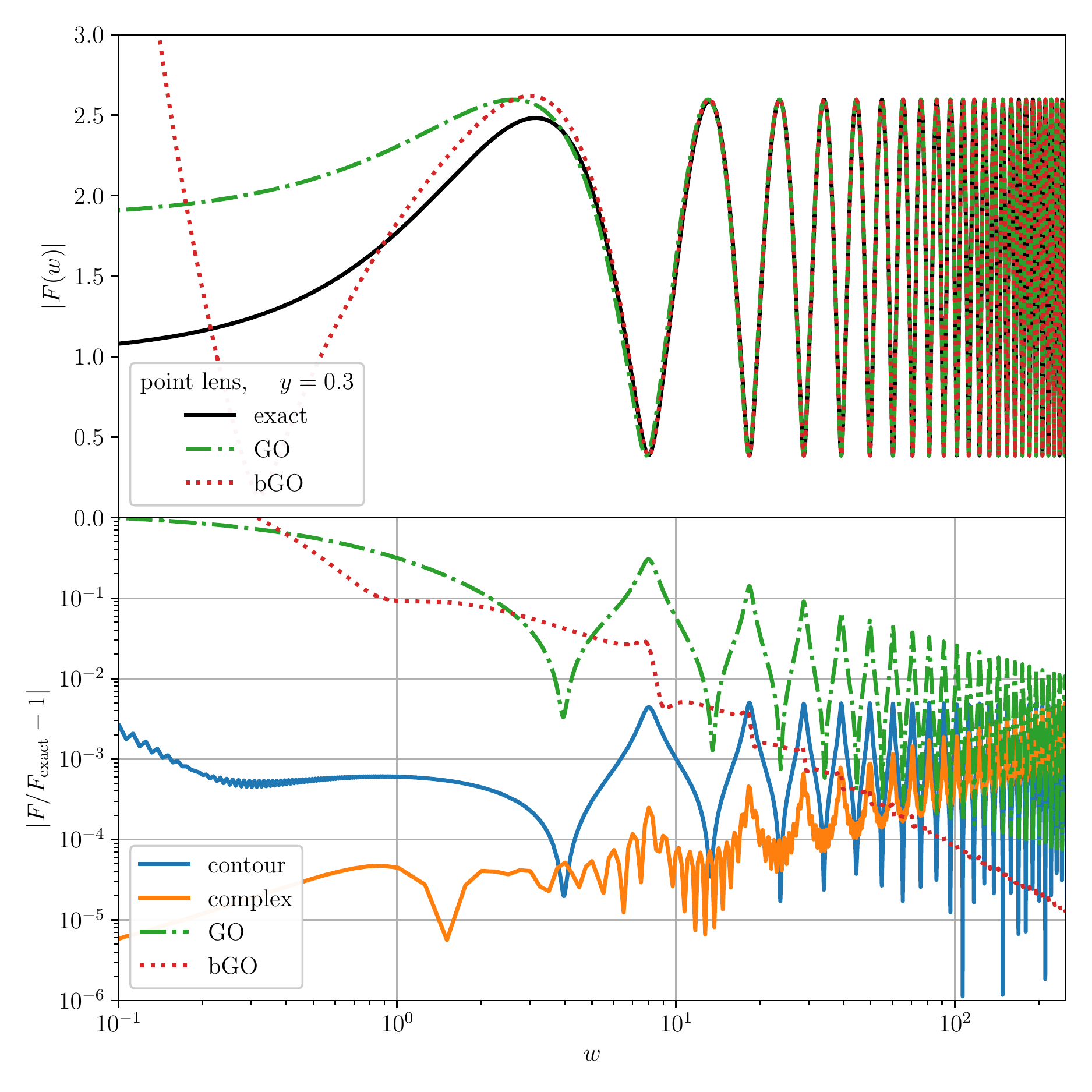}
\end{center}
\caption{
Methods comparison for a point lens ($y=0.3$).
\textbf{Top:} Absolute value of the amplification factor $|F(w)|$. The contour and complex method are not shown explicitly, as they overlap with the exact solution. 
\textbf{Bottom:} Differences relative to the exact solution.
The contour method (blue) always performs at the sub-percent level, while the complex-deformation method (orange) at low frequencies is around one order of magnitude better. The GO approximation (light green) converges very quickly, remaining sub-percent for $w>10$, whereas the bGO approximation (red) converges even faster.
}
\label{fig:method_comparison_point}
\end{figure}

We compare the result from the contour method of Sec.~\ref{sec:wo_contour} and the complex-deformation method of Sec.~\ref{sec:wo_complex_def} with Eq.~\eqref{eq:pt_lens_analytic} for $y=0.3$ in Fig.~\ref{fig:method_comparison_point}, where also the results of GO and bGO are reported. The contour method is most accurate in the intermediate-$w$ regime, up to $w \sim 10$, while the complex-deformation method remains good even at $w \sim 10^2$.

Let us now discuss the comparison with the contour method more in details, since it will be the primary method used in future applications, \cite{our_paper}. We will discuss the complex-deformation method's performance in the following subsection.
At low frequencies ($w\lesssim 0.1$) the numerical result from the contour method stops being accurate. This is related to the way the numerical calculation is performed. Indeed, the signal is obtained through a Fourier transform from the time domain signal $\tilde I(\tau)$. Therefore, the low-frequency errors are related to the numerical truncation of the integral in Eq.~\eqref{eq:contour_integrand} for large time delays, which depends on the windowing of $\tilde I(\tau)$. Higher precision can be achieved at low frequencies by extending the integral \eqref{eq:contour_integrand} to larger $\tau$s, at the expense of making the numerical evaluation slower.

In the opposite regime, for large $w$, we also lose accuracy. The appearance of error in this regime can be understood in the following way. At high $w$, the signal can be written as follows
\begin{equation}
    F(w) 
    =
    F_{\rm reg}^{(n_{\rm max})}(w) 
    + \sum_{J, n}^{n_{\rm max}} 
    \frac{c_n^{(J)}}{w^n}F_{\rm GO}^{J}(w)
    \;,
\end{equation}
where the sums are over the different images $J$ and the higher-order GO corrections that scale as $\sim w^{-n}$. Here for $n = 0$ we recover GO ($c_0^{(J)} = 1$) and for $n = 1$ we have instead bGO ($c_1^{(J)} = i \Delta_J$). The term $F_{\rm reg}^{(n_{\rm max})}(w)$ represents the regular WO contribution, not captured by the GO up to order $n_{\rm max}$.
We can notice that all the GO terms, when Fourier-transformed to the time domain $\tau$ give some ``singular'' features. In particular for $n = 0$, as we already discussed in Sec.~\ref{subsec:time_dom_reg}, we can have $\theta$-function discontinuities or $\log$ divergences in $\tilde I(\tau)$. As discussed there, there are also discontinuities/singularities on the $n$-th derivative of $\tilde I(\tau)$ with respect to $\tau$, for any $n$.
Due to finite numerical accuracy, such sharp features pollute the signal at arbitrary high $w$ when transformed back to frequency space. Having understood this,  we have a strategy for potential future improvements in the accuracy of our code. Indeed, these additional GO contributions could be subtracted before performing the inverse Fourier transform, in the same spirit of what is already done in the case of $n=0$. By performing this procedure up to $n = n_{\rm max}$, we expect the residuals against the full result to scale as $\sim 1/w^{1+n_{\rm max}}$. Of course, other sources of error might then become dominant.
We expect that these remarks also apply to other lens models.

\begin{figure}[t]
\begin{center}
  \includegraphics[width=0.98\columnwidth]{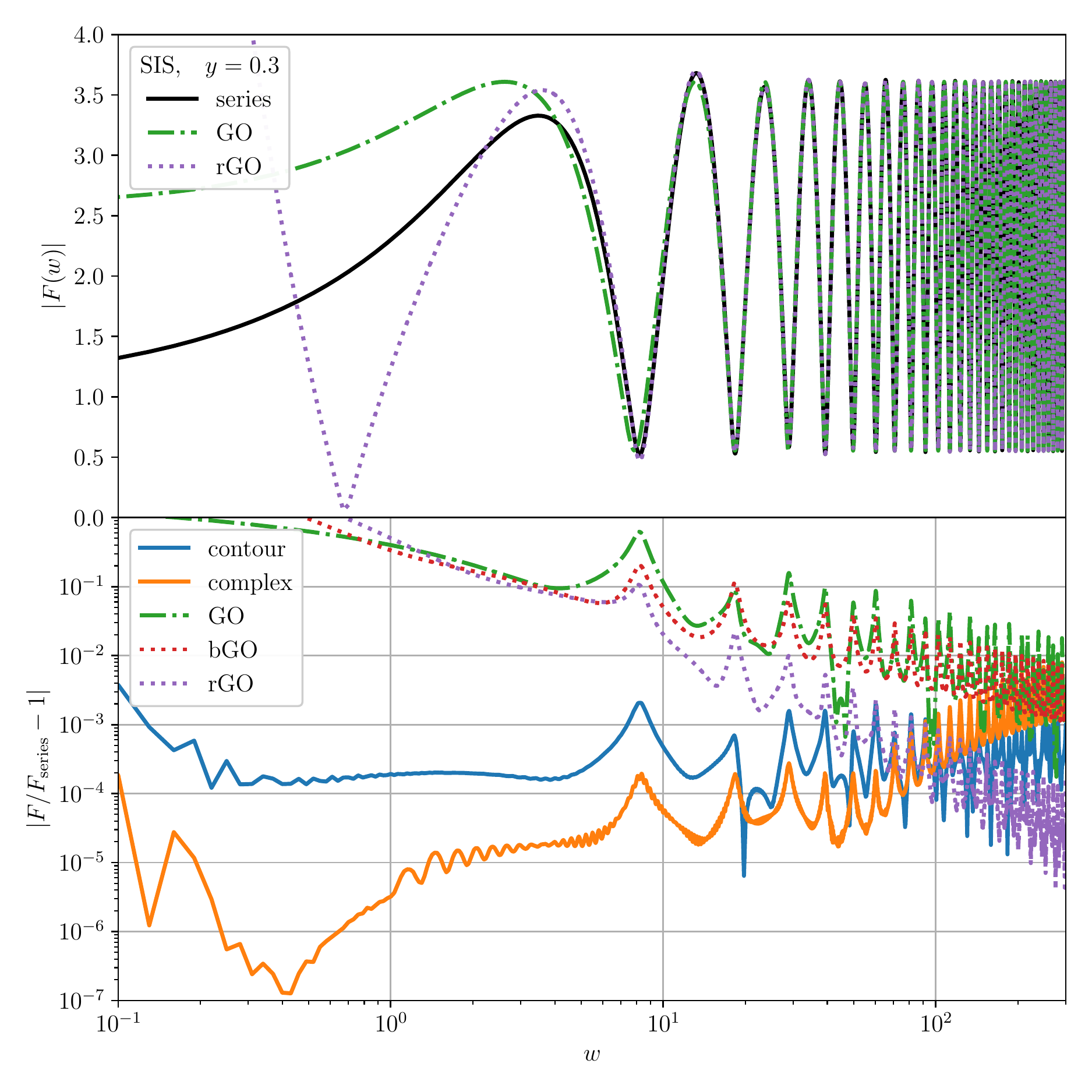}
\end{center}
\caption{Method comparison for the SIS  ($y=0.3$).
\textbf{Top:} Absolute value of the amplification factor $|F(w)|$. $F_{\rm series}$ is the series expansion of Eq.~\eqref{eq:sis_wo_exact}. The contour and complex method overlap with it and are not shown.
\textbf{Bottom:} difference relative to the series solution. In the WO regime both contour and complex-deformation methods perform below the per-mille level. The GO approximation converges to below percent level for $w \gtrsim 100$ and the bGO does not substantially improve convergence. The inclusion of the cusp information (rGO) gives a much faster convergence instead.}
\label{fig:method_comparison_sis}
\end{figure}

\begin{figure}
\begin{center}
  \includegraphics[width=\columnwidth]{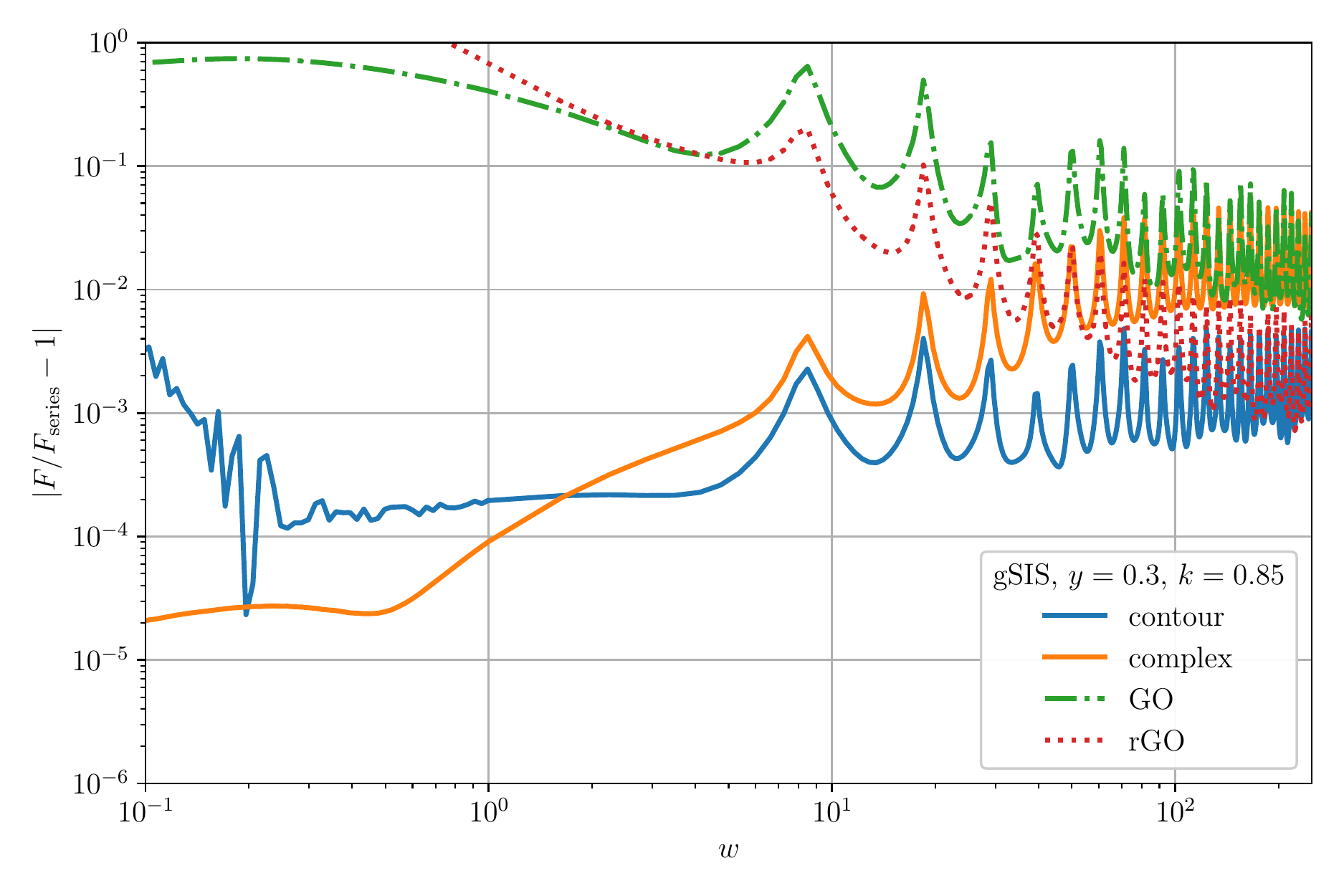}
\end{center}
\caption{Comparison between methods for a gSIS lens, where $F_{\rm series}$ is the series expansion of Eq.~\eqref{eq:gsis_wo_exact} truncated at some adequate $n$ as fo the SIS case in Fig.~\ref{fig:method_comparison_sis}. 
Instead of the usual bGO approximation, here we compare against the rGO approximation of Eq.~\eqref{eq:F_c_sol} (red).
The performance is overall very similar to the SIS case.
}
\label{fig:method_comparison_gsis}
\end{figure}

\begin{figure}
\begin{center}
  \includegraphics[width=\columnwidth]{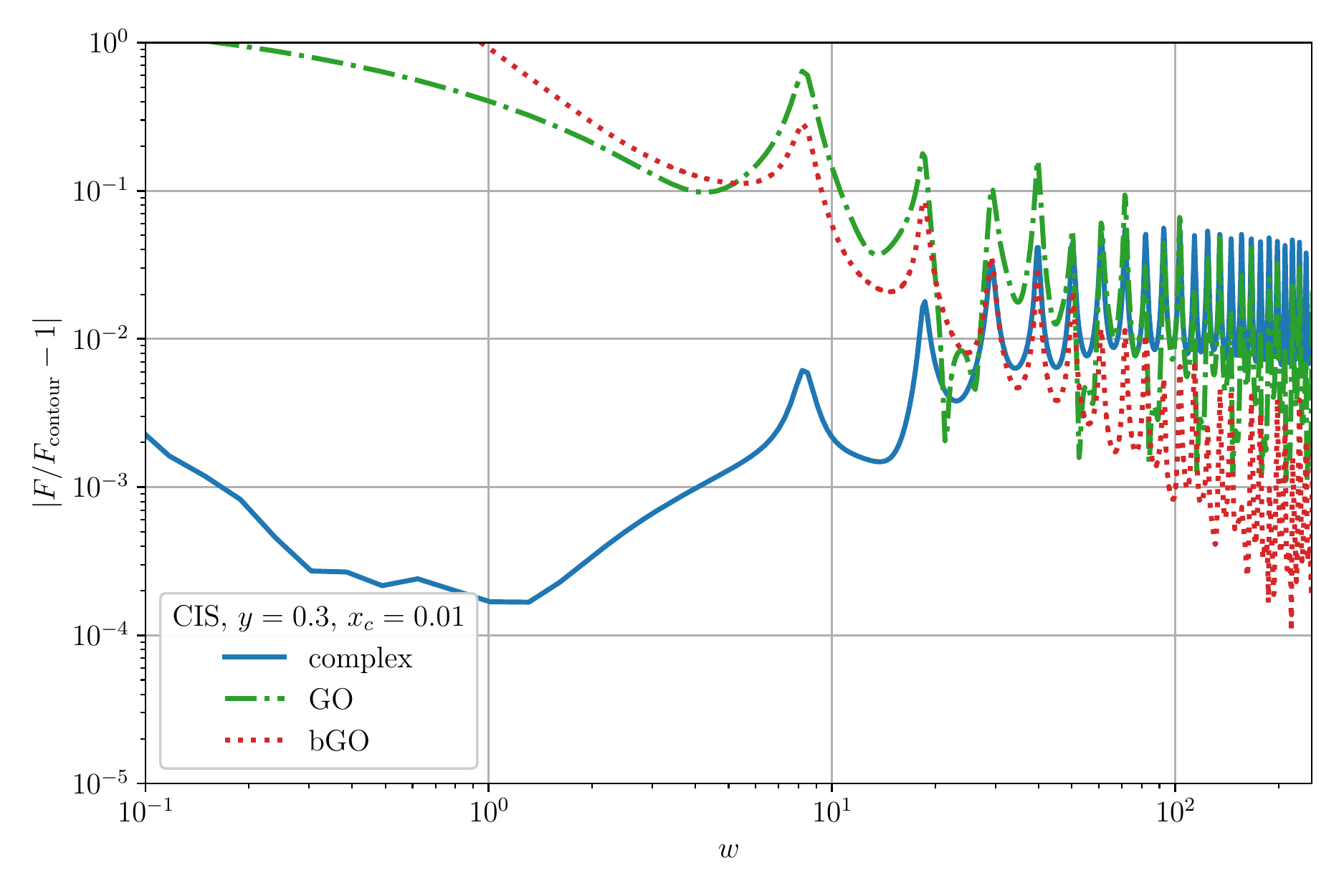}
\end{center}
\caption{
Comparison between methods for a CIS lens, where in this case the reference is taken to be the contour method $F_{\rm contour}$ (for CIS there is no explicit series representation for the amplification factor known to us). The performance is similar to the point-lens case, with GO and bGO methods becoming very reliable after $w>10$.
}
\label{fig:method_comparison_cis}
\end{figure}

\begin{figure}
\begin{center}
  \includegraphics[width=\columnwidth]{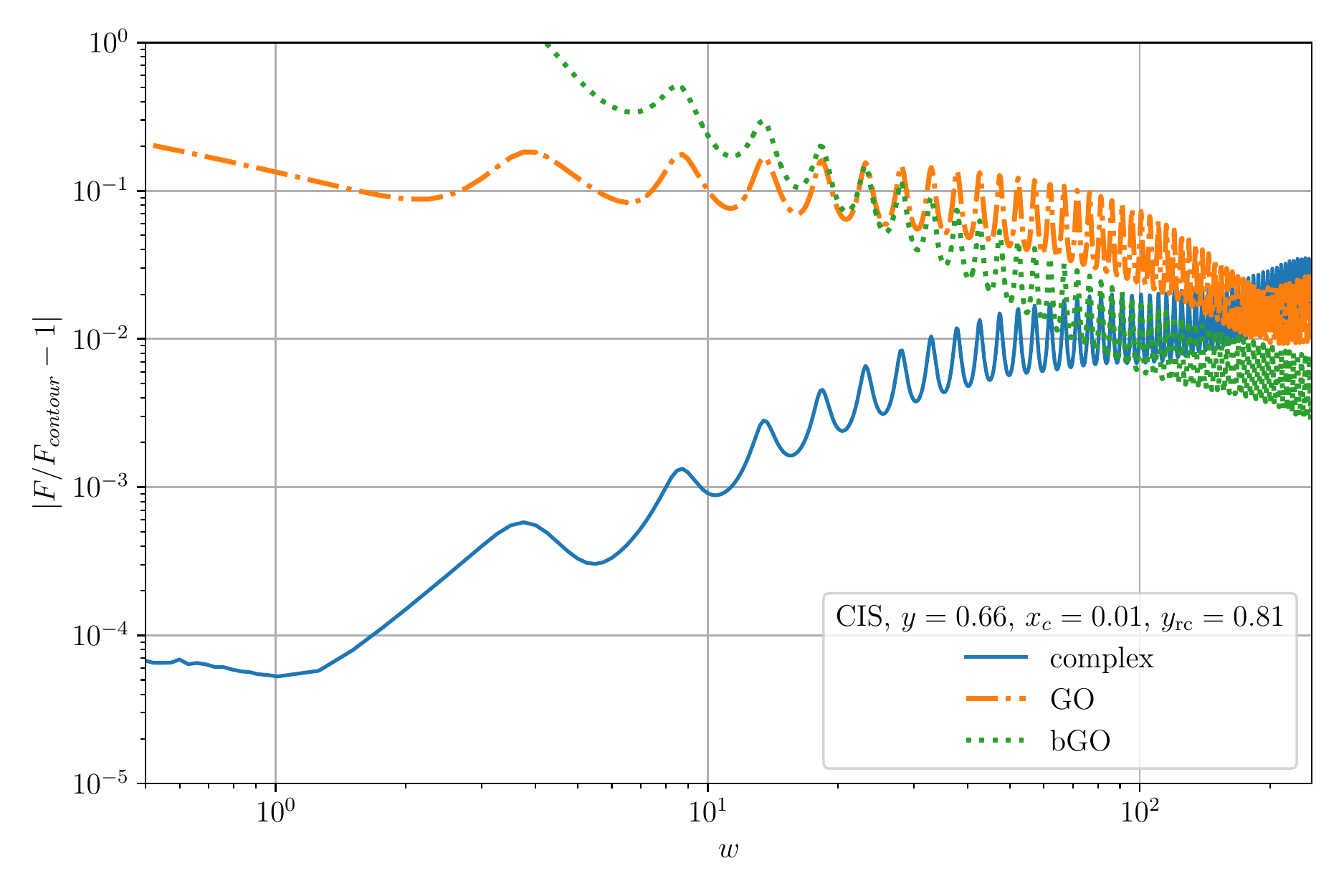}
\end{center}
\caption{
Comparison between methods for a CIS lens, as in Fig.~\ref{fig:method_comparison_cis} but with a larger impact parameter $y = 0.66$, closer to the caustic $y_{\rm rc} = 0.81$. 
In this case the accuracy degrades, but remains of order $1\%$ for $w < 100$.}
\label{fig:method_comparison_cis_yrc}
\end{figure}

\begin{figure}
\begin{center}
  \includegraphics[width=\columnwidth]{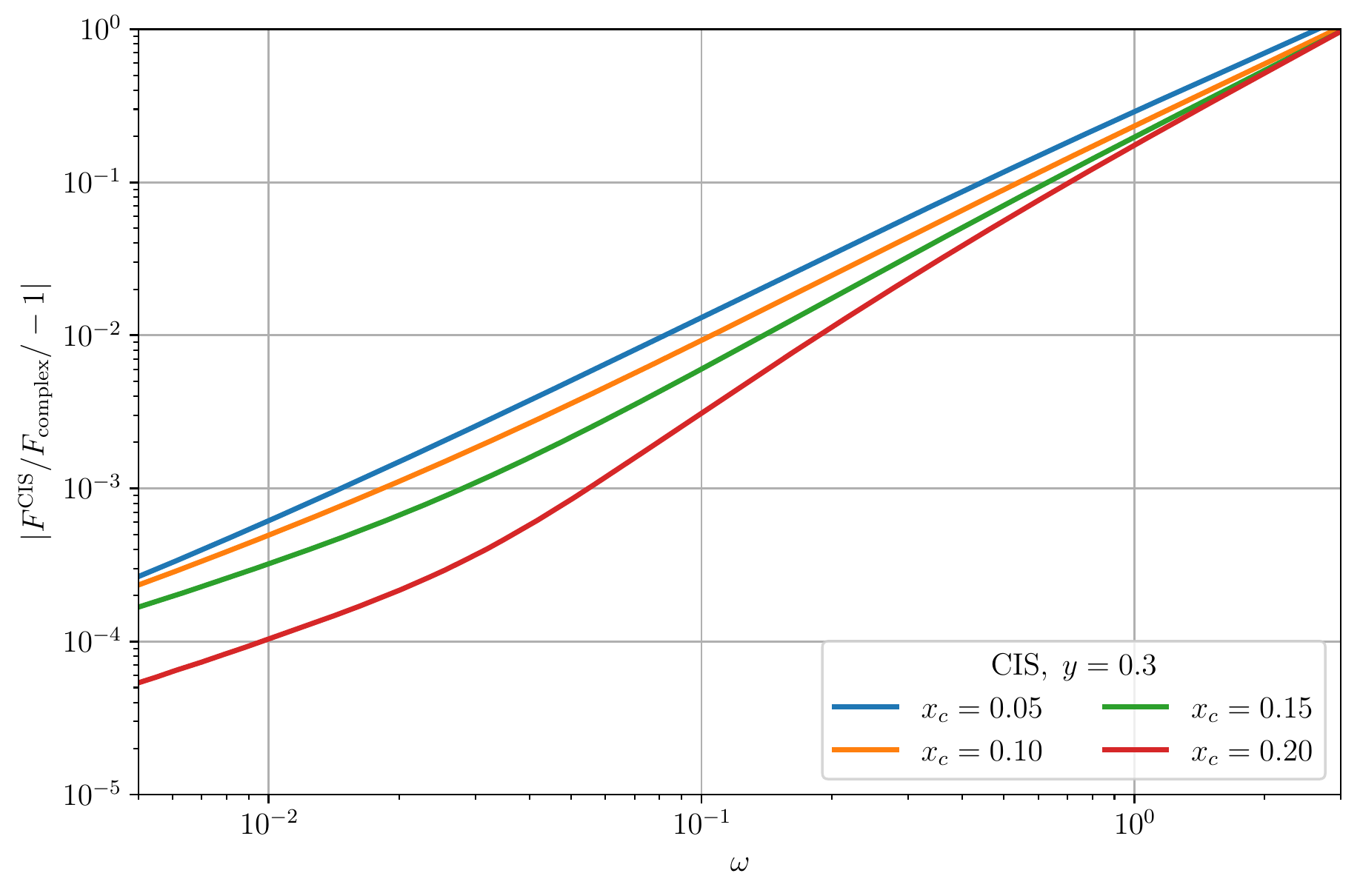}
\end{center}
\caption{
Comparison between the low-$w$ expansion $F^{\rm CIS}$ of Eq.~\eqref{eq:low_w_CIS_simpl} and the complex-deformation method of Sec.~\ref{sec:wo_complex_def} for a CIS lens with $y=0.3$. Different curves correspond to different values of the core size $x_c$. The analytic expansion reaches per-mille accuracy for $w \lesssim 10^{-2}$, while the expansion breaks down around $w \sim 1$, as expected.
}
\label{fig:low_w_comparison_cis}
\end{figure}

\subsection{Comparison between methods}\label{sec:compare_methods}

Here we compare the numerical results from the contour method and the complex-deformation method introduced respectively in Sec.~\ref{sec:wo_contour} and \ref{sec:wo_complex_def}. Again, we will consider the extended lenses described in Table \ref{tab:lenses_summary}.

For the SIS lens, as we have discussed in \ref{subsubsec:series_expansion}, a series representation for the integral is available and can be used for comparison with our numerical methods.
A comparison with the series Eq.~\eqref{eq:sis_wo_exact} is shown in Fig.~\ref{fig:method_comparison_sis}, again for $y=0.3$. We can see that the agreement for both our numerical methods is below the permille level in the range $10^{-1} \lesssim w \lesssim 10$.

At low frequencies the contour method starts to fail, as seen from the non-smooth curve for $F$. On the other hand, the complex-deformation method works best in this regime. It is more accurate since the (typically oscillatory) angular integral over $\theta$ in Eq.~\eqref{eq:complex_method_angular_int} is not particularly computationally demanding.\footnote{In the SIS case Fig.~\ref{fig:method_comparison_sis}, we start noticing a loss in accuracy when moving towards lower $w$s. This is because for small $w$ the integrand in Eq.~\eqref{eq:complex_deformation_C_integral} has support for large values of $|z|$ along the contour path $C_{\lambda}$. Since we truncate the path to a finite range, we effectively lose part of the integration domain. The problem is however straightforward to correct.
}
Going instead to higher $w$ reverses the situation: the complex-deformation method becomes slower and is outperformed by the contour one. The main reasons are the increasing oscillations in the angular integral and the fact that the computation has to be performed frequency-by-frequency (the contour method instead evaluates $F$ directly at all frequencies). 

In our code, the accuracy of the complex-deformation method can be improved by increasing the sampling of the path $C_{\lambda}$ and of the integral over $\theta$. This is at the cost of a longer evaluation time.

Let us compare different methods for computing the gSIS amplification factor.
We test the analytic result of Eq.~\eqref{eq:gsis_wo_exact} against the contour and complex-deformation methods. Additionally, we also check the improvement of the $\rm rGO$ approximation of Eq.~\eqref{eq:F_c_sol} against the GO-only result. The results are shown in Fig.~\ref{fig:method_comparison_gsis}. As we can see, the general results found for the SIS lens in Fig.~\ref{fig:method_comparison_sis} are preserved here, with similar orders of magnitude for the accuracy against the analytical result. Also, we can notice the slightly better agreement between the $\rm rGO$ compared with only GO (the bGO curve, not shown, gives a similar residual as the GO one).
With these results, we can establish that the contour and complex-deformation methods are accurate enough to study the gSIS lens. 

The final lens we consider is the CIS. In this case, as already discussed in Sec.~\ref{subsubsec:series_expansion}, a series representation is not available. Therefore, in Fig.~\ref{fig:method_comparison_cis} we compare the complex-deformation method against the contour method directly. Moreover, comparisons with the GO and bGO results are also shown.
As we can see, the two numerical methods have the lowest residuals around $w \sim 1$, whilst at lower and higher frequencies the results degrade. As for the other lenses, at low $w$ the contour method loses accuracy. At higher frequencies instead, the complex-deformation method become less reliable, making the residual bigger. We can notice that the trends are very similar to those of Fig.~\ref{fig:method_comparison_sis}, \ref{fig:method_comparison_gsis}, for the SIS and gSIS lenses. In comparing against GO and bGO we find similar features as for other lenses: as expected the residuals scale as $\sim 1/w$ and $1/w^2$ respectively. From this observation, we can argue that the contour method remains accurate below the percent level for $1 \lesssim w \lesssim 300$, at least. If that was not the case, the residual w.r.t.~GO and bGO would saturate at high frequencies.
For CIS there is no rGO curve, as the centre of the lens is regular and there is no cusp contribution.

We tested our results also at different impact parameters, obtaining similar agreements between methods. However, accuracy degrades as we approach a caustic where two images merge. 
Since images are closer in this situation, a lower precision is expected: resolving them requires higher resolution in the time delays (for the contour method) and higher resolution in the image plane (complex-deformation method).
In the case of the CIS, a caustic occurs at an impact parameter $y_{\rm rc}$ (see eq.~(64) in Ref.~\cite{our_paper} for the explicit expression). A comparison at a larger impact parameter is shown in Fig.~\ref{fig:method_comparison_cis_yrc}.

For SIS and gSIS lenses, the series representations (Eq.~\eqref{eq:sis_wo_exact} and \eqref{eq:gsis_wo_exact} respectively) reduce to the low-$w$ expansions (Eq.~\eqref{eq:low-w_SIS_low-y} and \eqref{eq:low-w_gSIS_low-y}) for small enough $w$. Therefore, the comparisons against the numerical methods in Fig.~\ref{fig:method_comparison_sis}, \ref{fig:method_comparison_gsis} implicitly show the goodness of this approximation. To perform a similar test for the CIS lens, we compare the low-$w$ expansion Eq.~\eqref{eq:low_w_CIS_simpl} against the complex-deformation method in Fig.~\ref{fig:low_w_comparison_cis}. The low-frequency approximation performs at sub-percent level for $w\lesssim 10^{-1}$, while per-mille accuracy is reached for $w\lesssim 10^{-2}$. Notice that the contour method would lose accuracy in this region.
These results have a mild dependence on the value of $x_c$, with more accurate results for larger $x_c$.

\subsection{Performance}\label{sec:compare_performance}

The contour method is by far the fastest computationally. Our implementation in \texttt{python} (optimized with \texttt{Numba}, after pre-compiling) computes $F(w)$ in the range $w_{\rm min} \sim 0.01, w_{\rm max} \sim 1000$ on a $12$-cores laptop (\texttt{i7-10750H} CPU) in $323\, {\rm ms}$, a range similar to the one used for Figs.~\ref{fig:method_comparison_point}-\ref{fig:method_comparison_cis} (the numbers refer to the point lens, the extended lenses require similar execution times).%
\footnote{This corresponds to $2^{17}\sim 10^5$ points in the FFT: increasing the number of points ($\propto w_{\rm max}/w_{\rm min}$) slows the computation due to the interpolation and FFT.}

The contour method is faster than computing the exact solution for the point lens, \eqref{eq:pt_lens_analytic}: it takes $5.1\,{\rm s}$ on the same machine sampling the same range of $w$ over $10^4$ points (a factor $1/10$ fewer values of $w$ than quoted above). A single evaluation of the exact solution \eqref{eq:pt_lens_analytic} takes $0.47\, {\rm ms}$ at $w\sim 0.01$ and $5.7\, {\rm ms}$ at $w\sim 1000$.

GO and bGO calculations are very inexpensive, requiring $12$  and $15\, {\rm ms}$  for the same lens, respectively. Because the contour method employs regularization, the GO amplification factor is already computed.

Complex deformation is efficient at low $w$ but each dimensionless frequency needs a separate computation. Moreover, the angular integral becomes highly oscillatory for large $w$ and needs to be sampled very finely to obtain the desired accuracy, at the cost of increasing the evaluation time considerably. 
With this method, one needs first to evolve the contours of integration (Panel B of Fig.~\ref{fig:complex_flow}) through the flow equation for different values of the angle $\theta$. This evaluation takes around $\sim 300\,{\rm ms}$ for a single angle and $100$ points in the $x$ integration path. Parallelization can be used to speed up the evaluation over multiple angles (for $\sim 50$ values of $\theta$ the overall evaluation takes $\sim {\rm few}\, {s}$). 
After this step, the evaluation of $F(w)$ at a single value of $w$ takes around $\sim 300\,{\rm ms}$. Most of the computational time is spent evaluating the integrand function over the $2{\rm D}$ domain of integration. Evaluation over multiple $w$s is then parallelized: this generally provides a factor $\sim 5$ gain in speed.
The situation worsens if $w$ is increased too much since the number of sampled angles $\theta$ needed for a precise evaluation has to increase (the timing will scale almost linearly with the number of angles).

Given the different execution times between both methods, employing the contour method in applications such as parameter-estimation for GWs is more convenient. The complex-contour method remains a valuable tool for the validation of our results, in particular at low frequencies. For instance, it allows us to test the low-$w$ behaviour derived in Sec.~\ref{subsec:wo_low} for different lens models.

\section{Conclusions}

In this work we have developed, implemented and validated methods to compute gravitational lensing predictions in the WO regime. 
We first outlined the general framework for WO computations and introduced several lenses that serve as examples for comparison (Sec.~\ref{sec:lensing_wo}). We described systematic expansions valid at high and low frequencies (Sec.~\ref{sec:lensing_expansions}). 
The geometric optics (GO) expansion is valid for $w\propto GM_{Lz}f\to \infty$ (Sec.~\ref{sec:lensing_go}). It gives the amplification as a sum over images, each of which carries a magnification, time delay and Morse phase. Finite frequency corrections $\propto 1/w$ can be systematically included, with terms that depend on derivatives of the lensing potential at the GO images. Similar $1/w$ corrections stem from non-smooth features of the lensing potential, which are not associated to GO images and vanish at sufficiently high frequency.

While GO is a local expansion around critical points, a low-frequency expansion highlights the dependence on the lens's global properties and asymptotic behaviour.
The $w\ll 1$ limit (Sec.~\ref{subsec:wo_low}) allows us to explore the deep WO regime and leads to simple expressions for our example lenses. 
The leading-order corrections depend on the asymptotic of the lensing potential, a trend clearly seen in the convergence towards free propagation $F \simeq 1$ as $w\to 0$:
\begin{itemize}
\item The point lens shows the fastest convergence, with $F-1\propto w + \mathcal{O}(w^2)$. This follows from $\psi \propto\log(x)$ being the slowest possible asymptotic growth of $|\psi|$.
\item Extended lenses approach the unlensed case more slowly, as $F-1\propto w^{k/2} + \mathcal{O}(w^{k/2+1})$ for $\psi\propto x^k$, and $k<2$ needed to keep the enclosed mass finite. 
\end{itemize}
This difference can be explained by the larger projected mass within a region of radius $\propto 1/w$, which dominates the diffraction integral at low frequencies. Hence, we expect that an extended but isolated lens (e.g.~truncated at finite radius) will recover the point lens convergence at sufficiently low $w$. 
The point-lens limit can not be recovered by the gSIS, as it requires $k \to 2$, leading to divergent mass.

While the low-frequency series can be computed, convergence requires including many terms, even at moderately high frequencies.
Intermediate frequencies depend on the properties of the lensing potential across the lens plane, which are more efficiently computed numerically.

We developed and presented two numerical methods to compute WO lensing.
The regularized contour-flow method (Sec.~\ref{sec:wo_contour}) solves the Fourier-transform of the diffraction integral (\ref{eq:lensing_wave optics}) by adaptively sampling equal-time contours of the Fermat potential. Subtracting the singular contributions and then adding their appropriate terms after transforming back to the frequency domain significantly reduces numerical noise at high frequencies.
The complex-deformation method (Sec.~\ref{sec:wo_complex_def}) analytically continues the integration variable in the complex plane. A well-defined process to flow the integration contour allows us to solve convergent, non-oscillatory integrals.

Both methods are complementary to each other. The method of contour flow allows a very fast computation of the amplification factor over a whole range of frequencies using FFT. The computation is efficient enough for GW parameter estimation with LVK data, although further optimization might be necessary to study complex lenses described by many parameters.
A main shortcoming of contour flow is that it requires knowing the initial and final conditions for each contour. While the endpoints are solutions to the lens equation, setting up a calculation for a complex setup with many images can become very involved. 
In contrast, complex deformation does not require knowledge of solutions to the lens equation. However, the method is costly, as each value of the $w$ needs to be computed independently. Moreover, high frequencies require a very fine sampling of the angular integral (this is not the case for contour flow, where one only needs to increase $N_{\rm FFT}$ to reach higher $w$).

We performed the first cross-validation of these numerical methods for different lensing potentials. We can achieve sub-percent accuracy over a broad range of frequencies with both methods. These results are optimal at intermediate frequencies, and can be matched to analytic expansions whenever the method fails or becomes too costly.
Besides comparing with the exact point-lens solution, we found excellent agreement with the series solution for the SIS and gSIS. Having two independent methods allows us to validate the predictions for lenses for which no systematic solution is known, as we demonstrated for the CIS. This will be important when considering more realistic and involved lens models.

Our main conclusions can be summarized as follows
\begin{enumerate}
  \item High- and low-frequency limits capture different properties of the lens. The large $w$ limit depends on the critical points and non-analytic features, while the low $w$ limit depends on the asymptotics of the lensing potential. Intermediate $w$ requires knowledge of the entire lensing potential.

 \item Predictions in any regime can be obtained efficiently by combining numerical methods at intermediate frequencies with analytical approximations at low/high frequency.
 
 \item We find sub-percent agreement to exact solutions, as well as between different methods in their regime of validity. Sub-percent accuracy holds for all the lenses we considered without fine-tuning the precision parameters.
 
 \item Numerical methods offer complementary advantages: contour flow gives frequency-dependent predictions very fast ($\lesssim 1\, {\rm s}$ on a laptop). Complex deformation does not require prior knowledge of solutions to the lens equation.
 
 \item These methods offer insights into WO phenomena. Complex deformation is particularly useful for understanding the low-frequency limit. The contour method gives a transparent interpretation of GO and bGO (as non-analytic features in the time-domain integral or its derivatives) and its splitting from other WO effects.
\end{enumerate}
Our methods provide a baseline for computing gravitational lensing predictions in the WO regime. However, many improvements are possible and will be desirable in the near future. A clear direction will be the extension to more complex gravitational lenses. This requires allowing the methods to work with tabulated values for the lensing potential (rather than closed-form expressions), streamlining steps such as finding the limits of contours or the sampling of integrals in terms of precision parameters. A goal of this program is to make the methods available and integrate them into a public software tool \cite{Birrer:2021wjl,Pagano:2020rwj}. We also envision further developments, such as including a robust computation of amplification factor derivatives to improve lens parameter inference. Our tools can also be generalized to other applications, such as plasma lensing \cite{Grillo:2018qjt,Wagner:2020ihx}, studying GW polarization effects \cite{Dalang:2021qhu,Oancea:2022szu,Oancea:2023hgu} and testing gravitational theories \cite{Ezquiaga:2020dao,Dalang:2020eaj,Chung:2021rcu,Goyal:2023uvm}.

The sub-percent accuracy we have demonstrated is sufficient for the near-term future of GW observations. In particular, it allows us to model lensed waveforms by LVK and even 3rd generation GW observatories, where we expect signal-to-noise (SNR) ratio $\lesssim 100$.
Higher accuracy might be possible by better adjusting the settings of our calculations, although refinements, such as including $1/w$ terms in the regularization scheme, will eventually become necessary. These improvements will eventually be required for analysing higher SNR sources, such as massive black hole binaries (${\rm SNR}\sim 10^3-10^4$) that will be observed by LISA and other space-borne detectors. Now and in the future, these methods will facilitate many applications, from searching WO effects in GW data to novel probes of the matter distribution in the universe and fundamental physics.


\acknowledgments{
It is a pleasure to thank S.~Savastano and H.~Villarrubia-Rojo for useful discussions.
L.D.~acknowledges the research grant support from the Alfred P.~Sloan Foundation (Award Number FG-2021-16495). M.H.-Y.C.~is a Croucher Scholar supported by the Croucher Foundation.
M.H.-Y.C.~is also supported by NSF Grants No.~AST-2006538, PHY-2207502, PHY-090003 and PHY-20043, and NASA Grants No.~19-ATP19-0051, 20-LPS20- 0011 and 21-ATP21-0010.
}

\appendix

\section{Derivatives of \texorpdfstring{$F(w)$}{F(w)} via Regularized Contour Flow} \label{sec:appendix_F_derivs}

Let us outline how one can evaluate numerical derivatives w.r.t.~the lensing parameters $\Theta_l$ of the WO amplification factor $F(w)$. This is relevant for instance in computing Fisher-matrix elements in a forecast for GW detectors.

We focus on the contour method since it is the most suitable for these applications.
Here, $F(w)$ and its time-domain counterpart $\tilde I(\tau)$ are split into singular and regular parts (recall Eqs.~\eqref{eq:contour_F_split}, \eqref{eq:contour_split}). 
We can evaluate derivatives on the singular part easily in the frequency domain (instead in the time domain one needs to take the difference of singular functions, which is problematic). 
The singular part only depends on the images, and we do not encounter problems when taking small numerical variations for the parameters $\Theta_l$.
On the other hand, derivatives of $F_{\rm reg}(w)$ can be computed in the time domain via finite differences and then Fourier transformed:
\begin{equation}\label{eq:forecast_bGO_derivs}
    \tilde I_{{\rm reg}, l}(\tau) 
    \simeq 
    \frac{\tilde I_{\rm reg}(\tau, \Theta_l+\epsilon)-\tilde I_{\rm reg}(\tau, \Theta_l)}{\epsilon}\,.    
\end{equation}
where $\tilde I_{{\rm reg}, l}(\tau) \equiv \partial \tilde I_{\rm reg}(\tau)/\partial \Theta_l$.
We chose $\epsilon$ so that $|\phi_J(\Theta_l+\epsilon)-\phi_J(\Theta_l)|<1/f_{\rm max}$ for all images $J$, so the Fermat potential of all images varies less than the grid spacing used in the Fourier transform. 
Despite working with the regularized integrand, differentiating promotes discontinuities in the derivatives (associated to bGO corrections) to discontinuities in the function. 
Numerical issues associated to these discontinuities can be ameliorated by further splitting as 
\begin{equation}
    \tilde I_{{\rm reg}, l} = \tilde I_{{\rm reg}, l}^{(2)} + \sum_J \Delta \tilde I^{(J)}_{{\rm sing}, l} \theta(\tau -\phi_J)\,,
\end{equation}
where 
$\Delta \tilde I^{(J)}_{{\rm sing}, l}$ 
is the discontinuity of the derivative of the regular part associated to the image $J$. Derivatives of the amplification factor $F(w)_{,l}$ are computed by adding the FFT of the regularized term $ \tilde I_{{\rm reg}, l}^{(2)}(\tau)$ and analytic expressions for the Fourier transform of the step functions, cf.~Eq.~\eqref{eq:contour_extrema_freqdomain}. 

\bibliography{gw_lensing}

\end{document}